\documentclass[apj]{emulateapj}
\usepackage{graphicx}
\usepackage{amsmath}
\usepackage{float}
\usepackage{epsf}
\usepackage[countmax]{subfloat}
\usepackage[usenames]{color}
\usepackage{natbib}
\usepackage[normalem]{ulem}

\renewcommand{\vec}[1]{\boldsymbol{#1}}

\shorttitle{Broad Emission-line Profiles from SBHBs in Circumbinary Disks}
\shortauthors{Nguyen et al.}

\begin{document}

\title{Emission Signatures from Sub-parsec Binary Supermassive Black Holes II: Effect of Accretion Disk Wind on Broad Emission Lines}

\author{Khai Nguyen\altaffilmark{1}, Tamara Bogdanovi\'c\altaffilmark{1}, Jessie C. Runnoe\altaffilmark{2}, Michael Eracleous\altaffilmark{3}, Steinn Sigurdsson\altaffilmark{3}, \& Todd Boroson\altaffilmark{4}}
\email{khainguyen, tamarab@gatech.edu}
\altaffiltext{1}{Center for Relativistic Astrophysics, School of Physics, Georgia Institute of Technology, 837 State St NW, Atlanta, GA 30332}
\altaffiltext{2}{Department of Astronomy, University of Michigan,1085 S. University Avenue, Ann Arbor, MI 48109}
\altaffiltext{3}{Department of Astronomy \& Astrophysics and Institute for Gravitation and the Cosmos, Pennsylvania State University, 525 Davey Lab, University Park, PA 16802}
\altaffiltext{4}{Las Cumbres Observatory Global Telescope Network, Goleta, CA 93117}

\begin{abstract}
We present an improved semi-analytic model for calculation of the broad optical emission-line signatures from sub-parsec supermassive black hole binaries (SBHBs) in circumbinary disks. The second-generation model improves upon the treatment of radiative transfer by taking into account the effect of the radiation driven accretion disk wind on the properties of the emission-line profiles. Analysis of 42.5 million modeled emission-line profiles shows that correlations between the profile properties and SBHB parameters identified in the first-generation model are preserved, indicating that their diagnostic power is not diminished. The profile shapes are a more sensitive measure of the binary orbital separation and the degree of alignment of the black hole mini-disks, and are less sensitive to the SBHB mass ratio and orbital eccentricity. We also find that modeled profile shapes are more compatible with the observed sample of SBHB candidates than with our control sample of regular AGNs. Furthermore, if the observed sample of SBHBs is made up of genuine binaries, it must include compact systems with comparable masses, and misaligned mini-disks. We note that the model described in this paper can be used to interpret the observed emission-line profiles once a sample of confirmed SBHBs is available but cannot be used to prove that the observed SBHB candidates are true binaries.
\end{abstract}


%1==============================================================================%

\section{Introduction}\label{sec:intro}

Gravitationally bound supermassive black hole binaries (SBHBs) are a product of galaxy mergers and progenitors of coalescing binaries, considered to be the prime sources for future space-based gravitational wave (GW) detectors. Expectations for detection of gravitational radiation from SBHBs have recently been raised by detections of their smaller counterparts, accomplished by the Laser Interferometer Gravitational Wave Observatory (LIGO), and by selection of the Laser Interferometer Space Antenna (LISA) for the large-class mission in the European Space Agency science program. In light of these developments, the rates at which SBHBs form and evolve to coalescence remain important open questions in black hole astrophysics. At the present time, the best avenue to address them is through electromagnetic observations and theoretical modeling.

This work is directly motivated by the ongoing spectroscopic searches, which have so far identified several dozen candidates for SBHBs with sub-parsec orbital separations. They rely on detection of the Doppler shift in the emission-line spectrum of an SBHB that arises as a consequence of orbital motion. In the context of the binary model, the Doppler shifted broad emission-lines are assumed to be associated with the gas accretion disks that are gravitationally bound to the individual SBHs \citep{gaskell83,gaskell96, eh94, bogdanovic09, shen10}. The main complication of this approach is that the Doppler-shift signature is not unique to SBHBs \citep[e.g.,][]{popovic12}. To address this ambiguity, more recent spectroscopic searches have been designed to monitor the offset of the broad emission-line profiles over multiple epochs and to target sources in which the modulation in the offset is consistent with orbital motion \citep{bon12, bon16, eracleous12, decarli13, ju13, liu13, shen13, runnoe15, runnoe17, li16, wang17}. The searches of this type, with yearly cadence of observations, are in principle sensitive to a subset of SBHBs with orbital separations $\lesssim10^4$ Schwarzschild radii. For every one SBHB in this range there should be hundreds of gravitationally bound systems with similar properties, at larger separations \citep{pflueger18}.

While the focus of the spectroscopic searches and other observational approaches has so far been on detection of SBHBs, an equally important question is: what can be learned once a statistically meaningful sample of binaries is available? In order to aid interpretation of spectroscopic searches for SBHBs, in a study preceding this one \citep[][hereafter Paper~I]{nguyen16}, we developed a semi-analytic model to describe the spectral emission-line signatures of SBHBs in circumbinary disks. In Paper~I, we have calculated a synthetic database of nearly 15 million broad optical emission-line profiles and explored the dependence of the profile shapes on characteristic properties of SBHBs. The main finding is that modeled profiles show distinct statistical properties as a function of the semimajor axis, mass ratio, eccentricity of the binary, and geometry of the circumbinary accretion flow. This suggests that the broad emission-line profiles associated with SBHBs can, in principle, be used to infer the distribution of these parameters and as such merit further investigation. 

The database of modeled profiles presented in Paper~I, however, could not be easily compared to the observed profiles of SBHB candidates. This is because it contains more diverse profile morphologies (reflected in $38-57$\% of profiles with multiple peaks), and on average broader profiles than the observed SBHB candidates or a non-binary population of active galactic nuclei (AGNs).  A majority of observed AGNs are characterized by single-peaked, broad emission-line profiles and only about 3\% have double-peaked profiles \citep{strateva03}. The published data on SBHB spectroscopic candidates so far do not show conclusive evidence for departure from this trend.

This discrepancy suggests that not all relevant physical phenomena have been captured by our first-generation model, a realization that has motivated an improved treatment of radiative transfer effects, presented in this work. Here, we investigate the effect of radiatively driven outflows from the accretion flow on the appearance of the emission-lines. Specifically, we calculate the emission-line profiles by taking into account propagation of line photons through the disk wind, re-evaluate the diagnostic power of broad emission-lines, and carry out a comparison of the observed and modeled profile samples. 

This paper is organized as follows: we describe improvements in the new model in \S\,\ref{sec:model}, present the results in \S\,\ref{sec:results}, provide comparison of the modeled and observed profiles in \S\,\ref{sec:implications}, discuss implications of our findings in \S\,\ref{sec:discussion}, and present our conclusions in \S\,\ref{sec:conclusions}.

%2==============================================================================%

\section{Description of the model}\label{sec:model}

%2.1==============================================================================%

\subsection{SBHB in the circumbinary disk}\label{sec:bmodel}

This work builds upon the model of a SBHB in a circumbinary disk presented in Paper~I.  We summarize the most important properties of this model here and refer the reader to Paper~I for more detail. The accretion flow around the SBHB is described as a set of three circular, Keplerian accretion disks: two mini-disks that are gravitationally bound to their individual SBHs, and a circumbinary disk. The three disks are modeled as independent broad line regions (BLRs), where the size of the two mini-disks, as well as the central opening in the circumbinary disk are constrained by the size of the binary orbit and are subject to tidal truncation by the binary SBH, as described in Paper~I.  

We assume that each of the two accreting SBHs can shine as AGN and illuminate their own mini-disk as well as the two other disks in the system. The bolometric luminosity of each AGN correlates with the accretion rate onto its SBH and photoionization by the AGN gives rise to the broad, low-ionization optical emission lines just like in ``ordinary" BLRs \citep{csd89,csd90}.  We establish the relative bolometric luminosities of the two AGNs based on the published measurements of accretion rates from simulations of SBHBs \citep[][]{al96,gr00,hayasaki07,roedig11,farris14}. The parametric model, adopted to describe the SBHB accretion rate ratio in this work, can be found in equation~3 of Paper~I. The emissivity of each disk can then be evaluated as a function of the accretion rate onto the SBHs and the disk size.

In this model the circumbinary disk is aligned with the SBHB orbit and we explore the effects of varying the relative orientation of the two mini-disks and the circumbinary disk. This setup is of interest because gravitational torques can cause precession while diffusive processes can align the SBH spins and the mini-disks axes with the orbital axis, and so the alignment of the system may evolve with binary separation \citep{miller13,hawley18}. Furthermore, the entire system is allowed to have arbitrary orientation relative to the observer's line of sight. This setup provides a variety of configurations in which the three disks are illuminated by the two AGN at different incidence angles. Note that disk misalignment can lead to the ``shielding" of one AGN by the mini-disk associated with the companion SBH, as seen from the perspective of a distant observer. We account for this effect when such configurations arise by allowing the blocked AGN to illuminate its own mini-disk and the circumbinary disk but not the mini-disk of the other SBH. On the other hand, we do not take into account the eclipse of one disk by another, which can arise in misaligned configurations.

We follow the emission-line profile calculations described by \citet{chen89}, \citet{chen89b} and \citet{eracleous95} to obtain an emission-line profile from each disk in the weak-field approximation. Contributions to the flux from the three disks are then summed into a resulting emission-line profile according to the calculation outlined in Paper~I. Using this approach, we create a database of profiles by drawing from a parameter space that describes physically motivated configurations of SBHBs and their associated circumbinary regions. In this work too, we focus on the H$\beta$ emission-line profiles, the second line of the hydrogen Balmer series, and note that this calculation is applicable to all permitted, low-ionization broad emission-line profiles.

%%%%%%%%%%%%%%%%%%%%%%%%%%%%%%%%%%%%%%%%%
%%%  TABLE 1
%%%%%%%%%%%%%%%%%%%%%%%%%%%%%%%%%%%%%%%%%
\begin{deluxetable}{ll}[H]  
%\tabletypesize{\scriptsize}
\tabletypesize{\normalsize}
\tablecolumns{4}
\tablewidth{0pt} 
\tablecaption{Parameters of the model}\label{table:parameters}
\tablehead{Parameter & Value}
\startdata
$q$ & 1 , 9/11 , 2/3 , 3/7 , 1/3 , 1/10 \\
$a/M$ & $5\times 10^3$ , $10^4$ , $5 \times 10^4$, $10^5$, $10^6$ \\
$e$ & 0.0 , 0.5 \\
$f$ & $0^{\circ}$, $72^{\circ}$, $144^{\circ}$, $216^{\circ}$, $288^{\circ}$ \\ 
$R_{\rm in1}/M_{1}$, $R_{\rm in2}/M_{2}$ & 500 , 1000\\
$R_{\rm out3}$ & $3a$ \\
$i$ & $5^{\circ}$, $55^{\circ}$, $105^{\circ}$, $155^{\circ}$\\
$\phi$  & $0^{\circ}$, $36^{\circ}$, $108^{\circ}$, $180^{\circ}$, $242^{\circ}$, $324^{\circ}$\\ 
$\theta_1$, $\theta_2$ & $0^{\circ}$, $30^{\circ}$, $60^{\circ}, 105^{\circ}$, $135^{\circ}$, $165^{\circ} $\\
$\phi_1$, $\phi_2$ & $0^{\circ}$, $25^{\circ}$, $60^{\circ}$, $185^{\circ}$, $210^{\circ}$, $235^{\circ}$\\
$h_1/M_{1}$, $h_2/M_{2}$ & 10\\
$\sigma/{\rm km\,s^{-1}}$ & 850 \\
$\tau_0$ &  0 ($10^{-4}$) , 0.1 , 1 , $10^2$ \\
$\lambda(R_{\rm in})$ & $10^{\circ}$ \\
$\eta$ & 1.0 \\
$\gamma$ & 1.2 \\ 
$b$ & 0.7 
\enddata
\tablecomments{$q$ -- SBHB mass ratio. $a$ -- Orbital semi-major axis. $e$ -- Orbital eccentricity. $f$ -- Orbital phase. $R_{{\rm in}, i}$, $R_{{\rm out},i}$ -- Inner and outer radius of the primary, secondary, or circumbinary disk. $i$ -- Inclination of the observer relative to the SBHB orbital angular momentum. $\phi$ -- Azimuthal orientation of the observer relative to the SBHB major axis. $\theta_i$, $\phi_i$ -- Inclination and azimuthal orientation of the primary and secondary mini-disk relative to the SBHB orbital angular momentum. $\sigma$ -- Turbulent velocity of the gas. $\tau_0$ -- Normalization of the disk wind optical depth. $\lambda$ -- Opening angle of the disk. $\eta$ -- Power law index in the description of gas density. $\gamma$, $b$ -- Parameters describing the wind velocity.}
\label{table:parameters}
\end{deluxetable}

Table~\ref{table:parameters} summarizes the parameter choices used to generate the second-generation database of profiles presented here. All but the final five parameters are identical to those used in Paper~I to create the first-generation database. We consider sub-parsec SBHBs with total mass $M= M_1 + M_2$ and mass ratios $q = M_2/M_1 \leq 1$, where $M_1$ and $M_2$ are the mass of the primary and secondary SBH, respectively. Note that we do not explicitly adopt a value for the SBHB mass, because the relevant properties and results of our calculation scale with this parameter (for e.g., any length scales and the monochromatic line flux defined in equation~\ref{eq:flux1}). The results are nevertheless valid for a range of masses that correspond to black holes powering regular, non-binary AGN (i.e., $\sim 10^6 - 10^9\,M_\odot$).

The binary orbits are characterized by a range of separations given by the orbital semi-major axes, $a$, expressed in units of $M\equiv GM/c^2 = 1.48\times10^{13}\,{\rm cm}\, (M/10^8\,M_\odot)$, where we use the binary mass as a measure of length in geometric units with $G = c = 1$. SBHBs are placed on either circular or eccentric orbits, encoded by the orbital eccentricity $e$. We choose five values to describe the orbital phase of SBHBs, $f$, which is measured from the orbital major axis to the position of the secondary SBH (see Paper~I for an illustration of geometry).

$R_{\rm {in}\it{i}}$ and $R_{\rm {out}\it{i}}$ represent the inner and outer radii of the BLRs associated with the three disks, where the subscript $i=1,2,3$ marks the primary, secondary, and circumbinary disk, respectively. In Table~\ref{table:parameters} the radii for the circumbinary disk are scaled to the total mass, while for the primary and secondary mini-disk they scale to the relevant SBH mass, for convenience. The outer radii of the two mini-disks range from approximately $700\,M$ to $5\times10^5\,M$ as a function of the binary orbital separation, eccentricity and the mass ratio \citep[see Paper~I and][]{paczynski77}. Furthermore, the size of the central cavity of the circumbinary disk is assumed to be $2a$ \citep{lp79, an05, macfadyen08}. Because they are not free parameters of the model, the sizes of the mini-disks and the radius of the central cavity of the circumbinary disk are not listed in Table~\ref{table:parameters}. 

The orientation of the observer relative to the SBHB orbit are given by the inclination and azimuthal angles, $i$ and $\phi$, respectively (see Figure~17 of Paper~I). $i$ describes the orientation of the observer's line of sight relative to the orbital angular momentum vector of the SBHB.  For example, $i = 0^\circ$ represents a clockwise binary seen face-on and values $i > 90^\circ$ represent counter-clockwise binaries. $\phi$ is measured in the binary orbital plane as the angle from the major axis to the projection of the observer's line of sight. 

Similarly, the angles $\theta_{\it{i}}$ and $\phi_{\it{i}}$ are used to describe the orientations of the mini disks with respect to the orbital angular momentum vector of the binary. For example, when $\theta_1 = \theta_2 = 0^\circ$, both mini-disks are coplanar with the SBHB orbit, and when $\theta_i > 90^\circ$, the gas in the mini-disks exhibits retrograde motion relative to the circumbinary disk. The azimuthal angles $\phi_i$ are measured in the binary orbital plane, from the orbital major axis to the projections of the mini-disk rotation axes. The orientation of the mini-disks is assumed to be ``frozen" over one orbital cycle of the SBHB (represented by a sequence of orbital phases) and they are not allowed to precess. In this model the circumbinary disk is assumed to always be coplanar and in co-rotation with the binary orbit.

We also assume that the central source of the continuum radiation associated with each SBH is compact and has spatial extent of $h_i = 10\,M_i$. Similarly, we describe the broadening of the emission-line profiles due to the random (turbulent) motion of the gas in each disk as $\sigma = 850\, {\rm km\, s^{-1}}$, a value appropriate for disk-like emitters \citep[e.g.,][]{eh94}.  The last five parameters in Table~\ref{table:parameters} encapsulate the properties of the accretion disk wind, which are discussed in the next section.

%2.2==============================================================================%

%%%%%%%%%%%%%%%%%%%%%%%%%%%%%%%%%%%%%%%%%
%%%  FIGURE 1
%%%%%%%%%%%%%%%%%%%%%%%%%%%%%%%%%%%%%%%%%
\begin{figure*}[t]
\centering
\includegraphics[width=0.85\textwidth, trim = {0 -5 0 -5}, clip=true]{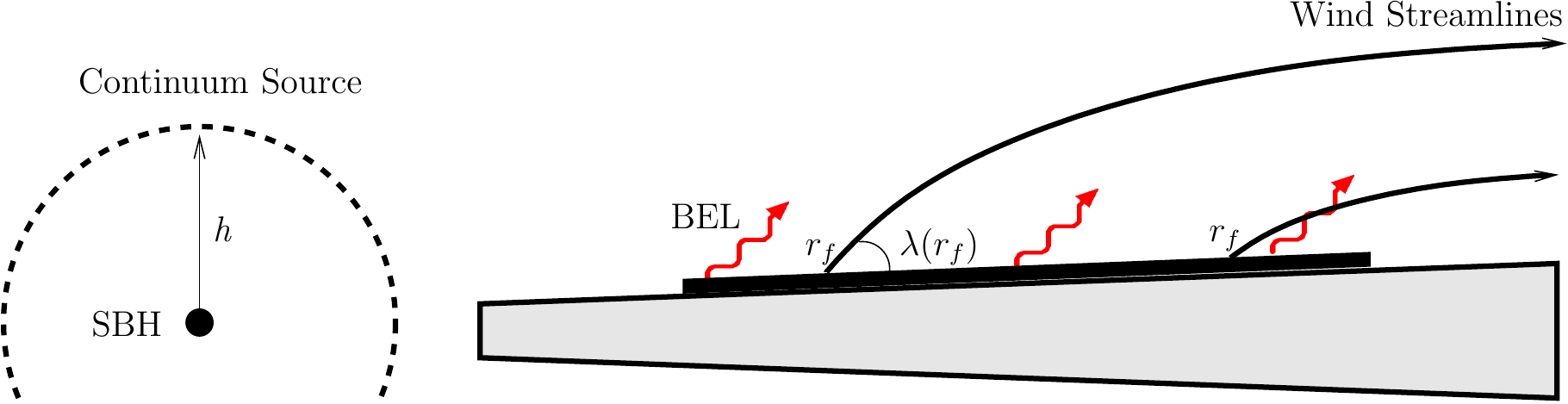} % l b r t
\caption{Illustration of the broad line region of a single SBH affected by the accretion disk wind, based on the model of \citet{cm96}. The compact source of continuum radiation (dashed circle) photoionizes the skin of an optically thick, geometrically thin disk giving rise to the low ionization, broad emission-line layer. Before escaping to infinity, some emission-line photons are absorbed by the accretion disk wind of finite optical depth, illustrated here as a set of streamlines lifting off the disk at the foot-point with radius $r_f$ at an angle $\lambda(r_f)$. Figure does not show the details of the inner accretion disk and is not to scale.}
\label{fig:geom}
\end{figure*}

\subsection{Disk wind model and parameter choices}\label{sec:dwmodel}

The theory that the accretion flow itself is the source of the broad emission lines is steadily gaining support in the AGN community. Studies of the response of the line profiles to changes in the flux continuum indicate that the motion of the gas in the H$\beta$ emitting BLRs of most AGNs is consistent with the thick disk and orbits that range from elliptical, to inflowing or outflowing trajectories \citep{wills86, bentz09,denney10,grier17,williams18}. Several works have demonstrated that disk models can be used to describe emission from BLRs of most AGNs when additional radiative transfer effects of the disk atmosphere on the emission-line profiles are accounted for \citep{cm96, mc97, flohic12, chajet13,chajet17, waters16, mangham17}. The origin of broad emission lines in the upper layer of an accretion disk, and the associated wind, is also compelling because the same wind scenario has been invoked to explain the broad, blueshifted absorption lines seen in the rest-frame UV spectra of a subset of AGNs \citep{mc95,proga04}. More recently, it has also been used to explain the existence of changing look AGNs \citep{macleod18}. 

%In addition to these, a number of authors have presented calculations of line profiles from an accretion disk wind applied to other astrophysical objects: for example, neutral hydrogen emission lines in young stars \citep[e.g.,][]{sim05} and high-ionization, UV resonance lines in cataclysmic variables \citep{shlosman93, vitello93, knigge95, knigge96}.

The origin of the line-driven wind in AGNs is in the inner accretion disk ($r\sim 10^{14}$\,cm for $\sim 10^8\,M_\odot$ SBH), where dense gas blocks the soft X-ray photons from the compact source of continuum radiation but transmits UV photons, which allows radiation pressure on resonance lines to accelerate the outflow to $\sim 0.1\,c$ \citep{mc97}. The wind extends to larger radii in the disk, where it affects the structure and kinematics of the BLR gas.  In this work, we explore this phenomenon in the context of the low-ionization H$\beta$ lines emerging from the BLRs surrounding SBHBs. We assume the H$\beta$ emission region to be a very thin layer on the surface of the outer accretion disk, which in AGNs extends from $\sim 10^{15}$\,cm to $\sim 10^{18}$\,cm  in radial direction. The emissivity above the emitting layer drops abruptly because hydrogen becomes highly ionized as a result of the steep decline in the density with height. The emissivity below the emitting layer drops sharply because the flux of ionizing photons from the central source at this depth is severely attenuated \citep[see photoionization calculation in Appendix~A of][]{flohic12}.

Figure~\ref{fig:geom} illustrates the geometry of the BLR around a single SBH in which emitted radiation is attenuated by the accretion disk wind. The radiation pressure lifts the gas from the surface of the disk and launches it along the wind streamlines, each of which is anchored to the disk at the foot-point with radius $r_f$, measured from the center of the disk in spherical coordinates. Each streamline makes a small angle, $\lambda$, relative to the disk which decreases as a function of radius \citep{cm96},
\begin{equation}\label{eq:wangle1} 
\lambda(r)=\lambda(R_{\rm in})\, \frac{R_{\rm in}}{r}
\end{equation} 
where $R_{\rm in}$ is the inner radius of each BLR (as defined in Table~\ref{table:parameters}) and $\lambda(R_{\rm in})=10^{\circ}$ is chosen for all three disks \citep{flohic12}. Before escaping to infinity, some low ionization emission-line photons are absorbed by a low density accretion disk wind. The wind is highly ionized and does not contribute significantly to the emission of low-ionization lines but has a finite optical depth in these lines, thus modifying the intensity and shape of the emitted profiles. 

Calculations of radiative transfer for outflows of this type are often carried out in the limit of large velocity gradient \citep[a.k.a., Sobolev approximation;][]{castor70, rybicki78, irons90, hamann93}. In this regime the photons that are not absorbed in the vicinity of the emission layer can escape to infinity, provided that the velocity of the wind projected onto the line of sight is monotonically increasing. Under such circumstances the photons do not encounter multiple regions along the line of sight where they can be absorbed.  Since accretion disk winds are expected to accelerate radially out \citep[e.g.,][and references therein]{proga00, proga04}, this condition is satisfied and the Sobolev approximation allows one to uncouple the absorption layer (marked as a black strip at the bottom of the wind streamlines in Figure~\ref{fig:geom}) from the rest of the wind. In this approximation, the characteristic thickness of the absorption layer is given by the Sobolev length, $\ell_S$. The probability that the low-ionization line photons escape the wind can then be estimated as a function of the local parameters in this layer
\begin{equation}\label{eq:escape1}
\beta_e(\vec{r},\,\vec{\hat{s}})=\frac{1-e^{-\tau(\vec{r},\,\vec{\hat{s}})}}{\tau(\vec{r},\,\vec{\hat{s}})} \,\,,
\end{equation}
where $\vec{r}$ marks the location from which the photon is emitted and the unit vector $\vec{\hat{s}}$ defines the direction of the observer's line of sight.  The line optical depth of the absorption layer, $\tau(\vec{r},\,\vec{\hat{s}})$, depends on the local mass density of neutral atoms, $\rho(\vec{r})$, opacity coefficient, $\varkappa(\vec{r})$, and turbulent velocity, $\sigma(\vec{r})$ 
\begin{equation}\label{eq:optd1}
\tau(\vec{r},\,\vec{\hat{s}})=\varkappa\,\rho\, \ell_S= \frac{\varkappa(\vec{r})\,\rho(\vec{r})\,\sigma(\vec{r})}{\left|\vec{\hat{s}}\cdot\vec{\Lambda}(\vec{r})\cdot\vec{\hat{s}}\right|} \,\,.
\end{equation}
Here, $\vec{\Lambda}$ is the wind velocity gradient tensor, which can be represented by its symmetric part (the rate of strain tensor) without  changing the resulting inner product, $Q \equiv \vec{\hat{s}}\cdot\vec{\Lambda}\cdot\vec{\hat{s}}$. Defined in this way, $Q$ is the velocity gradient of the wind along the line of sight. The model assumes constant $\varkappa$ and $\sigma$ within the thin absorption layer, and the density is expressed as a power law in radius, $\rho = \rho_0\, (r/M_i)^{-\eta}$, where $\rho_0$ is a normalization constant. Following \citet{flohic12}, we adopt  $\tau_0 = \varkappa\, \rho_0\, \sigma$, in which case equation \ref{eq:optd1} can be reduced to
\begin{equation}\label{eq:depth1}
\tau = \frac{\tau_0}{\left|Q\right|} \left(\frac{r}{M_i} \right)^{-\eta} \,\,,
\end{equation}
where $\tau \approx 5\tau_0$ ($7\tau_0$) represents the optical depth of the emission layer, along the direction perpendicular to the disk plane ($i=0^{\circ}$), at the inner edge of the BLR with $R_{{\rm in},i }= 500\,M_i$ ($1000\,M_i$), and assuming $\eta=1$. Note that equation~\ref{eq:optd1} implies that $Q$ must have units of inverse time in order for the optical depth, $\tau$, to be dimensionless. Keeping up with the formulation of equations in geometric units it follows that $Q$ and $\tau_0$ in equation~\ref{eq:depth1} (and hereafter) are expressed as dimensionless quantities in terms of $c^3/GM_i = M_i^{-1}$, and are properties that in this model decrease with the mass of the relevant SBH, or in the case of the circumbinary disk, binary mass. The details of this calculation, including the components of the $\vec{\Lambda}$ tensor and the final expression for $Q$, are shown in Appendix~\ref{sec:DW}. 

In this work, the emission line profiles are calculated for a range of optical depths, $\tau_0 = \left[10^{-4}, \, 10^{2} \right]$, as shown in Table~\ref{table:parameters}. In addition, Paper~I presents the emission-line profiles with $\tau_0 = 0$ (the ``no wind" scenario). Because the profiles calculated with $\tau_0 = 0$ and $10^{-4}$ are very similar, we use them interchangeably. We have also verified that profile shapes remain unchanged for $\tau_0>100$ and we do not explore the values of optical depth beyond this threshold. We further choose one value, $\eta=1$, to represent the radial dependance of the wind density, after verifying that the impact of this parameter on the profile shapes is relatively weak. See Appendix~\ref{sec:DW} for a more detailed discussion about these parameter choices.

An additional ingredient necessary for this calculation is the description of the poloidal component of the wind velocity along a given streamline
\begin{equation} \label{eq:velr}
v_p(r)=v_{\infty} \left(1- b\, \frac{r_f}{r}\right)^{\gamma} \,\,.
\end{equation}
At the launching point on the surface of the disk (i.e., at the foot-point of the streamline), we assume that the wind velocity is comparable to the Keplerian velocity in the disk, $v_p(r_f) = (r_f/M_i)^{-1/2}$, resulting in a total speed of the wind close to the escape speed from the SBH. A choice of $b=0.7$ and $\gamma=1.2$, adopted here, then implies that the wind accelerates to the terminal velocity $v_{\infty} \approx 4.7\, (r_f/M_i)^{-1/2}$, which corresponds to $v_{\infty}\approx 0.2\,$c for the launching point at $r_f=500\,M_i$ (see Appendix~\ref{sec:DWvelocity}). The value of $\gamma=1.2$ is consistent with the values inferred from observations, which range from $1.06$ to $1.3$, from quasars to Seyferts, respectively \citep{mc95}.

We assume that the disk wind driven by each AGN extends over the entire surface of its BLR. It is not clear however whether these outflows can extend from the mini-disks into the circumbinary disk, especially in configurations in which the disks are not co-planar. In order to examine this effect we calculate profiles for three different disk wind configurations, described below.

\begin{itemize}
\item NW -- This is the ``no wind" configuration presented in Paper~I, which corresponds to a disk wind model with $\tau_0 = 0$. In this limit, the probability of escape for the line photons, defined in equation~\ref{eq:escape1}, defaults to $\beta_e = 1$. The NW database contains nearly 15 million modeled profiles -- 2,545,200 realizations of SBHBs on circular orbits and 12,273,000 on eccentric orbits.
\item 2DW -- A disk wind develops only along the two SBH mini-disks and not in the circumbinary disk. In this setup, we calculate profiles from SBHB systems on circular orbits and with three different values of optical depth, $\tau_0=0.1$, $1$, $100$. The 2DW database contains about 7.5 million profiles -- 2,545,200 realizations for every value of the optical depth.
\item 3DW -- A disk wind is present in all three disks. The circumbinary disk has an accretion disk wind which is radial and axisymmetric, as if driven by a single, central AGN. Here, we calculate profiles from circular SBHBs with $\tau_0=0.1$, $1$, $100$, and from eccentric SBHBs with $\tau_0=1$. The 3DW database contains about 20 million profiles -- 2,545,200 realizations of circular SBHBs for every value of optical depth and 12,273,000 realizations of eccentric SBHBs.
\end{itemize}

Combined together, the entire database contains about 42.5 million profiles that correspond to the same number of SBHB configurations. Note that the nominal number of simulations per disk wind configuration is determined as a product of the number of parameter choices. From these simulations we eliminate the ones in which the orientation of any disk with respect to the observer is close to edge on (between $80^{\circ}$ and $100^{\circ}$). We do this to prevent the breakdown of the weak-field approximation, used in calculation of the photon Doppler shifts (see equation~17 in Paper~I). This selection criterion eliminates scenarios in which the impact parameter of the line-of-sight photons flying over a SMBH becomes too small (i.e., $\lesssim 100\,M_i$). Such photons experience significant gravitational redshift and gravitational bending of their trajectory. Because this happens in a small fraction of all SBHB configurations that we consider, we do not perform calculations of these effects in the strong field regime.

We describe different models by labels that encapsulate the description of the SBHB orbit, the disk wind model, and the optical depth normalization value. For example C-2DW-100 represents a set of profiles for SBHBs on circular orbits and a disk wind characterized by $\tau_0=100$ in both SBH mini-disks but not in the circumbinary disk. In contrast, E-3DW-1 refers to a database of profiles calculated for SBHBs on eccentric orbits, where $\tau_0=1$ in all three disks.

%3==============================================================================%

\section{Results}\label{sec:results}

%3.1 ==============================================================================%

\subsection{The effect of wind optical depth on profile peaks}\label{sec:NOP}
 
\begin{figure}[t]
\centering
\includegraphics[width=0.47\textwidth, clip=true]{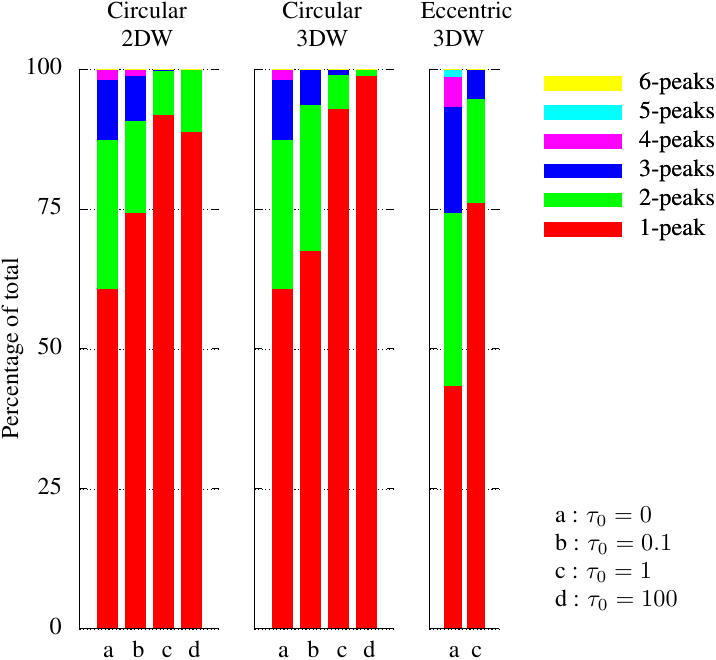}
\caption{Percentage of single- and multi-peaked profiles in different models calculated for SBHBs on circular and eccentric orbits. 
In all models the number of single-peaked profiles increases with the optical depth in the disk wind. SBHBs on eccentric orbits tend to have profiles with a higher number of peaks relative to SBHBs on circular orbits. Each column corresponds to a different value of the optical depth, $\tau_0$: (a) 0, (b) 0.1, (c) 1 and (d) 100. Different colors mark profiles with one to six peaks.}
\label{fig:NOP}
\end{figure}

In the next two sections we report properties of the emission-line profiles produced in this work and compare them to the first-generation model, in which radiative transfer through the disk wind was not accounted for. Figure~\ref{fig:NOP} provides a comparison between the first- and second-generation models in terms of the number of peaks that characterize their emission-line profiles. In the absence of a disk wind, each BLR can give rise to a double-peaked broad emission-line profile. Therefore, the combinations of three BLRs can produce a composite broad profile with up to six distinct peaks, depending on the relative motion of the BLRs with respect to the observer. This is indeed reported in Paper~I and shown in Figure~\ref{fig:NOP} in column ``a" for each model. These columns reflect the profile demographics in the NW model, which corresponds to $\tau_0 = 0$, whereas cases ``b", ``c" and ``d" correspond to the increasing optical depth in the disk wind. 

Figure~\ref{fig:NOP} shows that in the absence of the accretion disk wind, the fraction of multi-peaked profiles reaches 38\% for SBHBs on circular orbits and 57\% for eccentric SBHBs. As reported in Paper~I, SBHBs on eccentric orbits tend to have profiles with a higher number of peaks relative to the circular binaries with the same semimajor axis because the eccentric SBHBs sample a wider range of orbital velocities, allowing for a larger wavelength offset of individual components in the composite profile. 

The most important trend captured by Figure~\ref{fig:NOP} is the increase in the percentage of the single-peaked profiles with the wind optical depth in each model. For example, for SBHBs on circular orbits in the 3DW model the number of single-peaked profiles increases from $62\%$ in the $\tau_0 = 0$ case to $98\%$ in the $\tau_0 = 100$ case. The remaining 2\% of profiles in $\tau_0 = 100$ case are double-peaked profiles, and there are no profiles with three or more peaks.  A similar trend can also be found in the eccentric 3DW model, where even a moderate optical depth ($\tau_0 = 1$) eliminates complex profiles with more than three peaks.

As mentioned in the introduction, the multi-peaked profiles produced by the first-generation model do not reflect the average properties of the observed sample of SBHB candidates, or the general population of AGNs, the majority of which tend to have broad but single-peaked profiles. This discrepancy provided the main motivation for further development of the model, which now includes radiative transfer through the disk wind. Figure~\ref{fig:NOP} shows that this development results in model profiles that are consistent with observations given some appropriate value of $\tau_0$.  For example, only about 3\% of observed AGNs exhibit the double-peaked broad optical emission-line profiles \citep{strateva03}. If sub-parsec SBHBs in circumbinary disks follow a similar trend, our models indicate that their accretion disks must have outflows with substantial optical depths ($\tau_0 > 1$).  

%%%%%%%%%%%%%%%%%%%%%%%%%%%%%%%%%%%%%%%%%
%%%  FIGURE 2
%%%%%%%%%%%%%%%%%%%%%%%%%%%%%%%%%%%%%%%%%

%%%%%%%%%%%%%%%%%%%%%%%%%%%%%%%%%%%%%%%%%
%%%  FIGURE 3
%%%%%%%%%%%%%%%%%%%%%%%%%%%%%%%%%%%%%%%%%
\begin{figure*}[t]
\centering
\includegraphics[width=0.95\textwidth, clip=true]{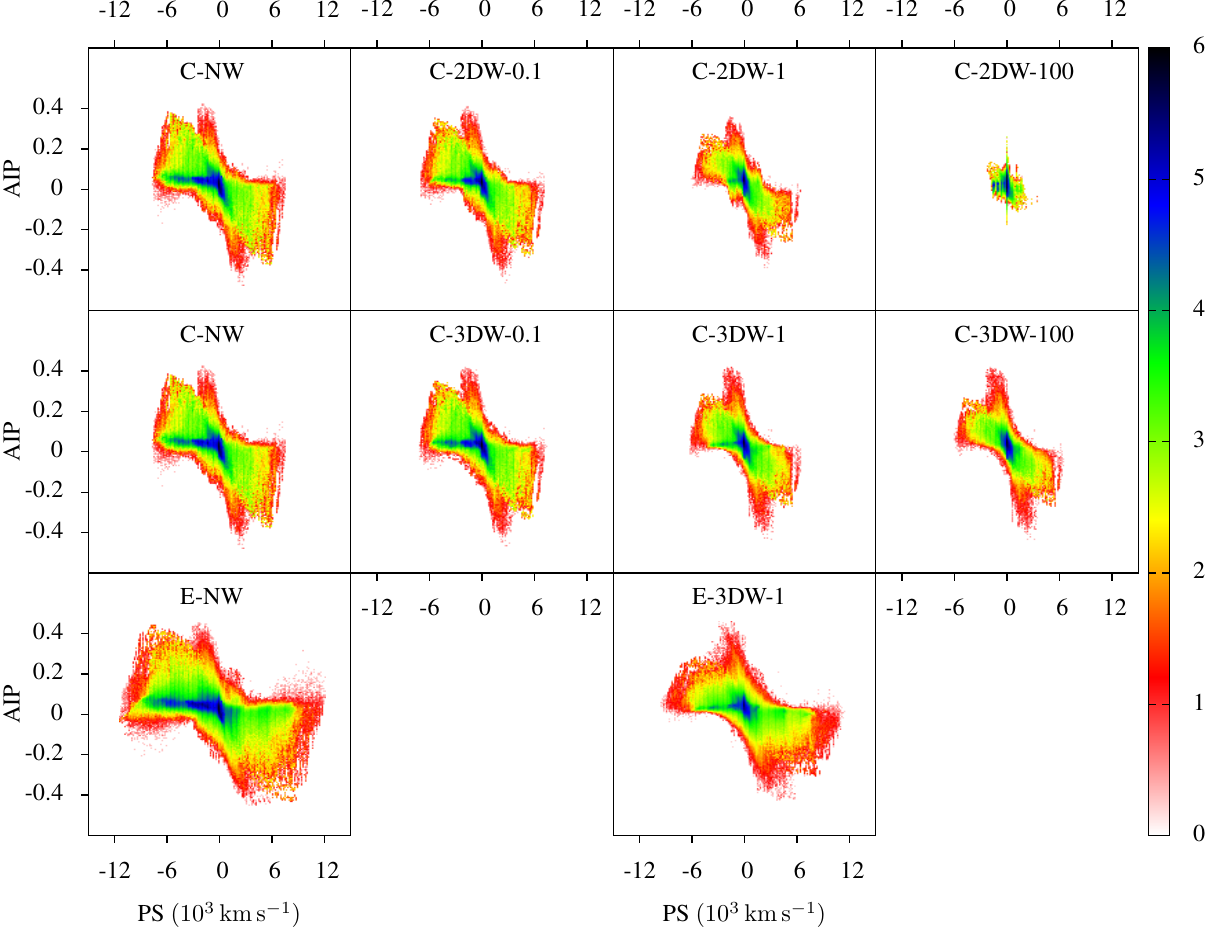}
\caption{Maps of Pearson skewness coefficient vs. peak shift (AIP-PS) for profiles associated with models of SBHBs on circular and eccentric orbits, different wind configurations, and wind optical depths. The first two rows depict profile distributions in the models with SBHBs on circular orbits where there is no absorption of the broad emission-line photons by the disk wind (NW models), the absorption occurs in the two mini-disks (2DW), or in all three disks (3DW). Third row shows the same for the models with eccentric SBHBs. In all rows the wind optical depth increases from left to right, from $\tau_0=0$ (NW models) to 0.1, 1, and 100 for DW models. Color bar indicates the density of profiles (i.e., the number of profiles in each area element) plotted on a log scale.}
\label{fig:AIPPSa}
\end{figure*}

Estimating the fraction of double-peaked profiles in the datasets obtained from the observational searches for sub-parsec SBHBs is however non-trivial. For example, the sample of candidates selected for spectroscopic monitoring by \citet{eracleous12} and \citet{runnoe15,runnoe17}\footnote{We refer to it as the E12 search hereafter.} includes only the sources which were in the first epoch of observations characterized by single-peaked emission-line profiles.  Further analysis of this sample has shown that after the subtraction from the H$\beta$ complex of the narrow H$\beta$ line and [O\,\textsc{iii}] doublet, 17/88 (or about 20\%) of SBHB candidates exhibit apparent double-peaked broad H$\beta$ line profiles. Because the subtraction of the narrow line components is not unique, it introduces an uncertainty that can make the resulting broad line appear double-peaked. We therefore conclude that $\lesssim20\%$ of the SBHB candidates in the E12 sample have line profiles that are truly double-peaked. If all of these are shown to be genuine SBHBs, this would require them to have disk winds with $\tau_0 \gtrsim 0.1$.

It is interesting to note that in the 2DW model for SBHBs on circular orbits, the number of single-peaked profiles increases for optical depth $\tau_0 = 0.1$ but then levels off for $\tau_0 = 1$ and 100. In this setup, the attenuation of emitted radiation in the disk wind is only present in the two mini-disks but not in the circumbinary disk. Once the optical depth in the mini-disks becomes substantial, their emission is significantly attenuated relative to the circumbinary disk. This is because re-emission of the H$\beta$ line photons absorbed in the wind is unlikely, given that a hydrogen atom in $n=4$ energy state can reach the ground state via several different radiative decay channels, can be radiatively or collisionally ionized, and collisionally excited or de-excited. While some of these processes may result in re-emission of the H$\beta$ photons, their numbers should be considerably smaller relative to the number of absorbed ones. Consequently, the circumbinary disk remains the dominant contributor to the composite broad emission-line profile. Therefore, the number of multi-peaked profiles is in the high optical depth limit determined by the number of double-peaked profiles contributed by the circumbinary disks.

%%%%%%%%%%%%%%%%%%%%%%%%%%%%%%%%%%%%%%%%%
%%%  FIGURE 4
%%%%%%%%%%%%%%%%%%%%%%%%%%%%%%%%%%%%%%%%%
\begin{figure*}[t]
\centering
\includegraphics[width=0.68\textwidth, clip=true]{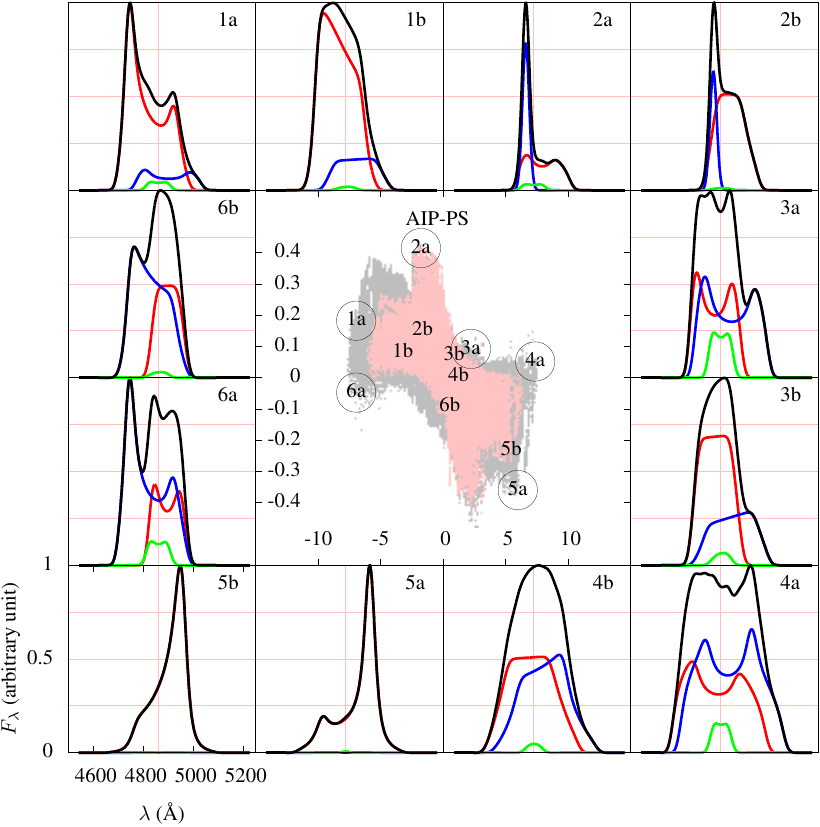}
\caption{Appearance of individual emission-line profiles in C-NW and C-3DW-100 models. {\it Central panel:} Following Fig.~\ref{fig:AIPPSa}, gray color represents the profile distribution for $\tau_0 = 0$, while the pink overlay marks the $\tau_0 = 100$ case. Markers ``a" and ``b" trace the location of profiles calculated for the same SBHB configurations with zero and high optical depth, respectively. {\it Surrounding panels} illustrate the appearance of profiles associated with markers in the central panel. In all cases, higher optical depth in the disk wind gives rise to more symmetric profiles with a smaller number of peaks. Flux is shown in arbitrary units against wavelength marked on the bottom $x$-axis. Pink vertical line at 4860.09$\AA$ marks the rest wavelength of the H$\beta$ emission line. For all profiles, total flux (black line) is a sum of components contributed by the primary (red), secondary (blue), and circumbinary disk (green).}
\label{fig:AIPPSb}
\end{figure*}
%
%%%%%%%%%%%%%%%%%%%%%%%%%%%%%%%%%%%%%%%%%
%%%  TABLE 2
%%%%%%%%%%%%%%%%%%%%%%%%%%%%%%%%%%%%%%%%%
%
\begin{deluxetable*}{cccccccccccccc}[H]  
\tabletypesize{\normalsize}
\tablecolumns{14}
\tablewidth{0pt} 
\tablecaption{Physical Parameters of Profiles Shown in Figure 4} \label{table:f4params}
\tablehead{Profile & $a$ & $q$ & $e$ & $R_{\rm in1}$ & $R_{\rm in2}$ & $i$ & $\phi$ & $\theta_1$ & $\phi_1$ & $\theta_2$ & $\phi_2$  & $F_{\lambda, {\rm b}}^{\rm max} /F_{\lambda, {\rm a}}^{\rm max}$ }
\startdata
1 & 5000 &   0.818182 &   0.0 &    1000 &       1000 &        105 &          0 &         60 &         60 &        105 &        235 &   0.0011 \\ 
2 & 5000 &   0.818182 &   0.0 &     500 &       1000 &        105 &          0 &        105 &        185 &        105 &          0 &   0.0013 \\ 
3 & 5000 &   0.818182 &   0.0 &    1000 &       1000 &        105 &          0 &         60 &          0 &        135 &        235 &   0.0017 \\ 
4 & 5000 &   0.666667 &   0.0 &     500 &        500 &         55 &          0 &        165 &         60 &         30 &        235 &   0.0029 \\ 
5 & 5000 &   0.100000 &   0.0 &     500 &       1000 &          5 &          0 &        105 &          0 &        105 &        235 &   0.0014 \\ 
6 & 5000 &   1.000000 &   0.0 &    1000 &       1000 &        105 &          0 &        135 &          0 &        165 &         60 &   0.0018 
\enddata
\tablecomments{$F_{\lambda, {\rm b}}^{\rm max} /F_{\lambda, {\rm a}}^{\rm max}$ -- the ratio between the maximum value of the flux for the attenuated profile in panel b (C-3DW-100 model), and the corresponding profile in panel a (C-NW model), before they were normalized to 1.}
\label{table:profiles}
\end{deluxetable*}

%3.2==============================================================================%

\subsection{Characteristic features of the modeled emission-line profiles}\label{sec:char}

Following the approach laid out in Paper~I we analyze the trends in the modeled group of profiles by characterizing their shapes in terms of several commonly used distribution functions.  These include the Pearson skewness coefficient (AIP), full width at half and quarter maximum (FWHM and FWQM), peak shift (PS), and centroid shift (CS), defined in equations 7--13 of Paper~I. We choose these properties among other distribution functions because they provide robust measures of the dominant features in the bulk of the profile. We avoid profile shape parameters that rely on the wings on the line profiles as these are significantly affected by the noise present in the observed spectra. For more detailed analysis of the impact of the noise on spectral measurements and statistical distribution functions of profiles see Appendix~C in Paper~I and Appendix in \citet{runnoe15}.

In Figure~\ref{fig:AIPPSa} we visualize the distribution of profiles in two-dimensional maps of AIP versus PS values calculated for all models, including SBHBs on circular and eccentric orbits, different wind configurations, and wind optical depths. The color marks the number density of profiles on a logarithmic scale and indicates which portions of the parameter space are favored by the modeled profiles. By definition, positive values of AIP indicate profiles skewed toward short wavelengths, i.e., blue-leaning profiles, and negative values indicate red-leaning profiles. Similarly, negative values of PS indicate that the highest (or the only) peak of the profile is blueshifted with respect to the rest wavelength of the emission line and vice versa.

Inspection of the panels in Figure~\ref{fig:AIPPSa} reveals that in all models a significant fraction of profiles are fairly symmetric (AIP $\approx 0$) and likely to exhibit their highest peak at wavelengths shorter than the rest wavelength (PS $< 0\, {\rm km\, s^{-1}}$). The latter is a consequence of relativistic Doppler boosting, which for each individual disk preferentially boosts the blue shoulder of its emission-line profile, creating an effect that is also noticeable in the composite profile. Another feature worth noting is that in all but one sample (C-2DW-100), the profiles that exhibit the strongest peak at shorter wavelengths are also preferentially blue-leaning and vice versa. This is of interest because this trend is also present in the sample of SBHB candidates observed as a part of the E12 search \citep[see Figure~18 in][and Section~\ref{sec:comparison} in this paper for further discussion]{runnoe15}.

%%%%%%%%%%%%%%%%%%%%%%%%%%%%%%%%%%%%%%%%%
%%%  FIGURE 5
%%%%%%%%%%%%%%%%%%%%%%%%%%%%%%%%%%%%%%%%%
\begin{figure*}[t]
\centering
\includegraphics[width=0.85\textwidth, clip=true]{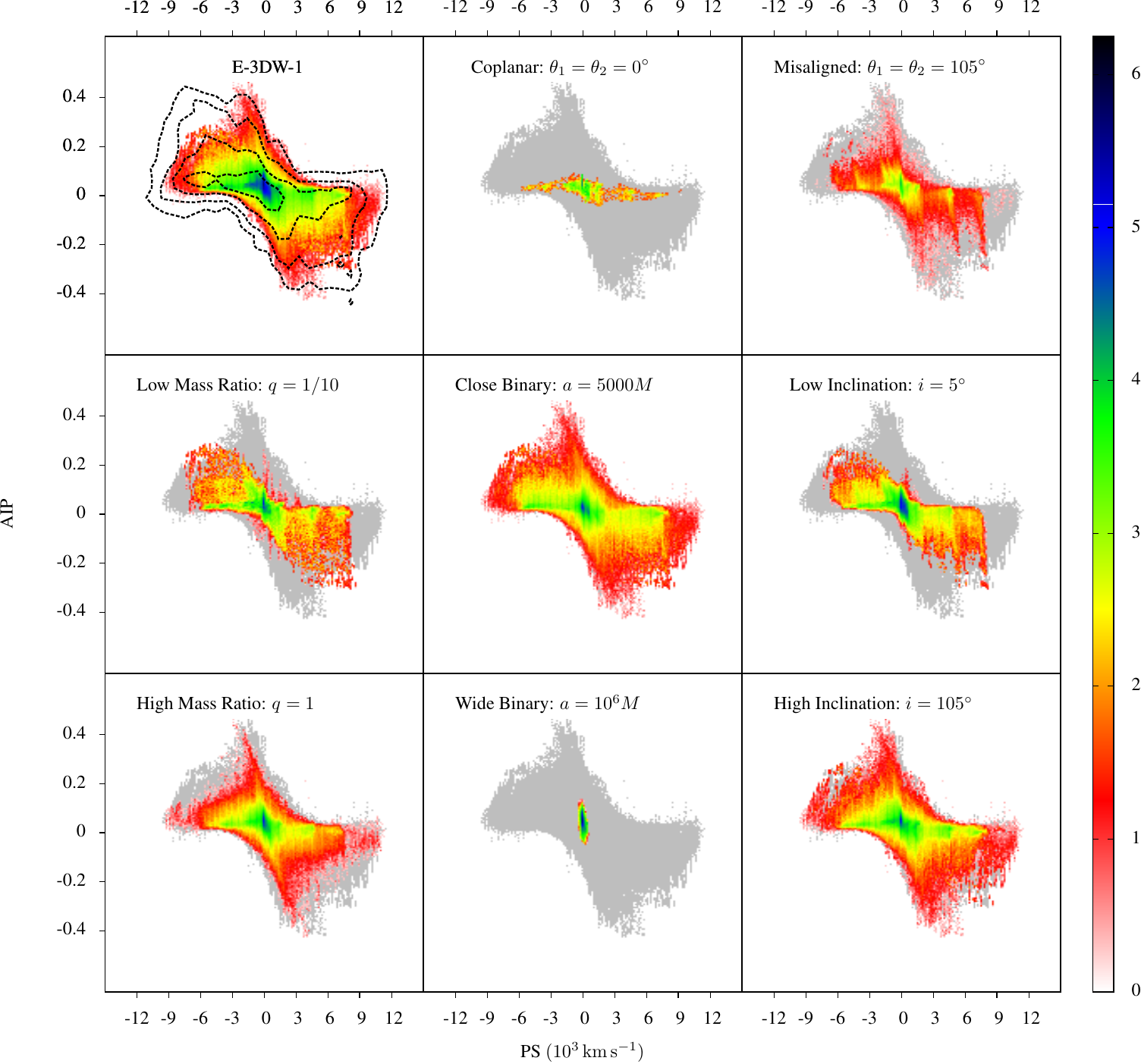}
\caption{AIP-PS map for profiles associated with the eccentric SBHB model E-3DW-1 (top left). Remaining panels display the distribution of profiles as a function of the alignment of the triple disk system, SBHB mass ratio, orbital separation, and inclination of the observer relative to the binary orbit. Color bar indicates the density of profiles (i.e., the number of profiles in each area element) plotted on the $\log$ scale. Grey color outlines the footprint of the entire distribution shown in top left panel. Black dashed contours in the top left panel are drawn in increments of one, from 0.5 (outermost) to 3.5 (innermost). They correspond to the AIP-PS maps in the NW model (as shown in Figure~7 of Paper I) and are included here for easier visual comparison.}
\label{fig:catAIPPS}
\end{figure*}

%%%%%%%%%%%%%%%%%%%%%%%%%%%%%%%%%%%%%%%%%
%%%  FIGURE 6
%%%%%%%%%%%%%%%%%%%%%%%%%%%%%%%%%%%%%%%%%
\begin{figure*}[t]
\centering
\includegraphics[width=0.85\textwidth, clip=true]{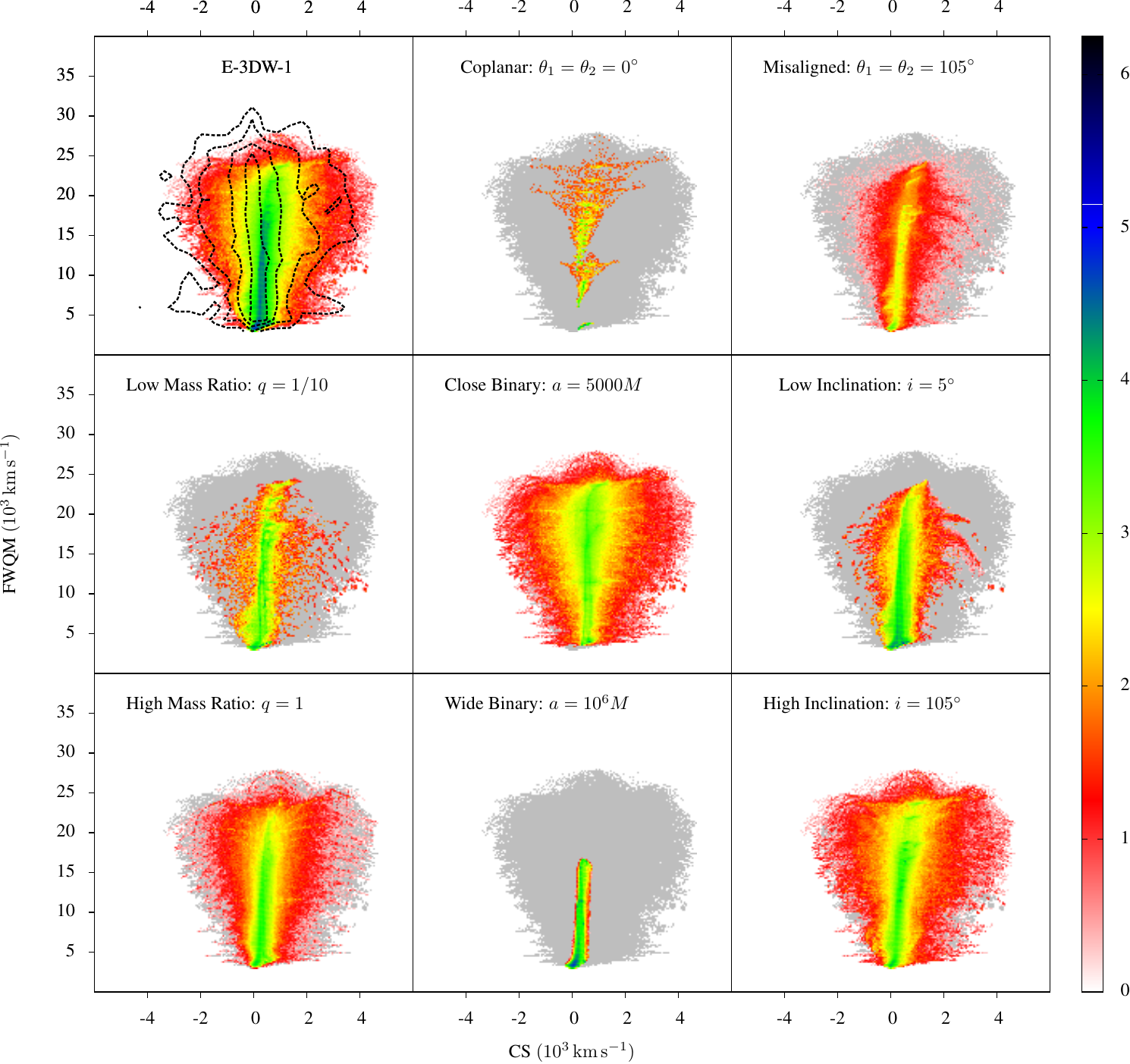}
\caption{Maps of full width at quarter maximum vs. centroid shift (FWQM-CS) for profiles associated with eccentric SBHB systems in model E-3DW-1. Black dashed contours in the top left panel correspond to the FWQM-CS maps in the NW model (as shown in Figure~11 of Paper I). The map legend is the same as in Figure~\ref{fig:catAIPPS}.}
\label{fig:catFWQMCS}
\end{figure*}

%%%%%%%%%%%%%%%%%%%%%%%%%%%%%%%%%%%%%%%%%
%%%  FIGURE 7
%%%%%%%%%%%%%%%%%%%%%%%%%%%%%%%%%%%%%%%%%
\begin{figure*}[t]
\centering
\includegraphics[width=0.85\textwidth, clip=true]{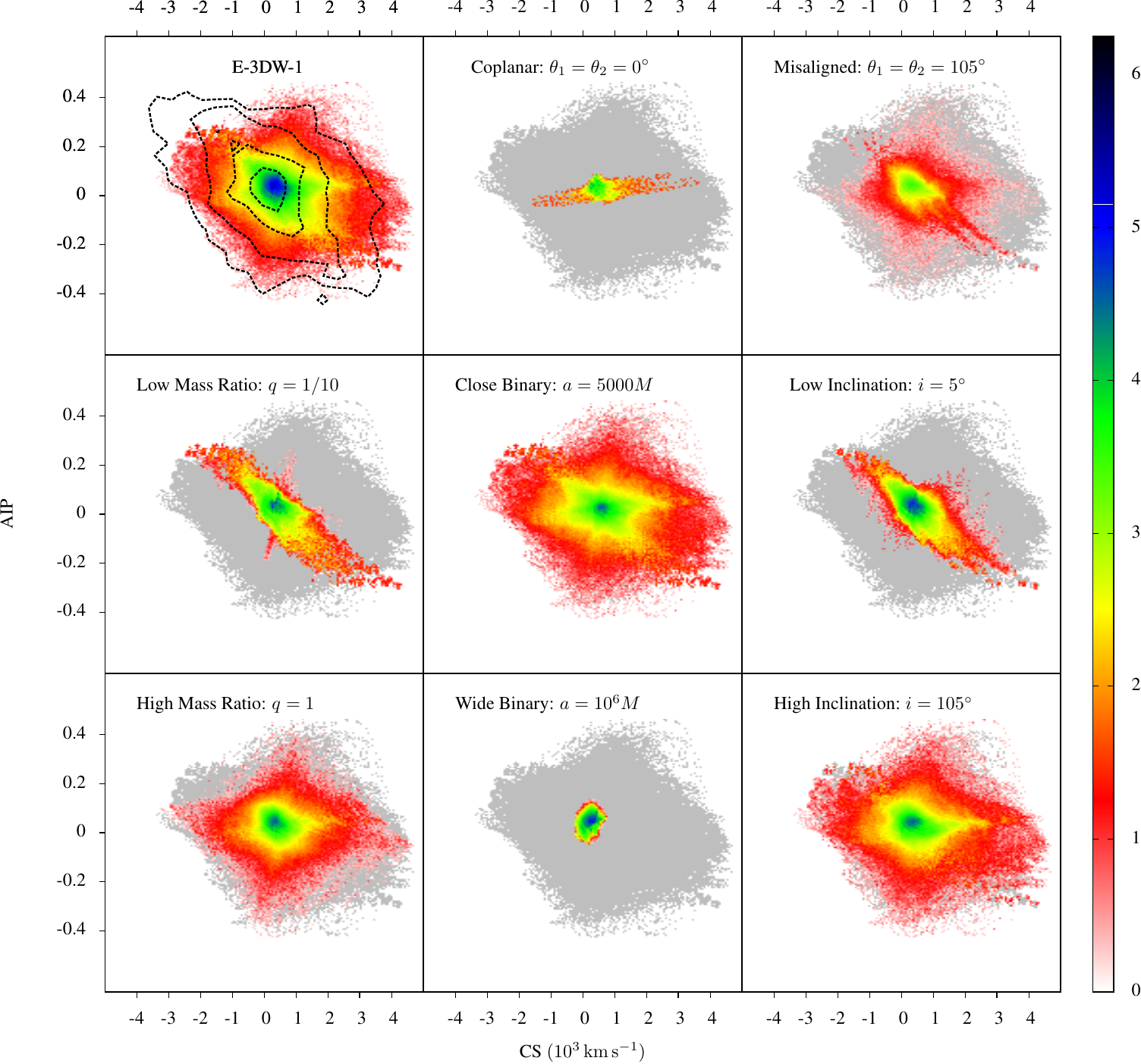}
\caption{Maps of Pearson skewness coefficient vs. centroid shift (AIP-CS) for profiles associated with eccentric SBHB systems in model E-3DW-1. Black dashed contours in the top left panel correspond to the AIP-CS maps in the NW model (as shown in Figure~15 of Paper I). The map legend is the same as in Figure~\ref{fig:catAIPPS}.}
\label{fig:catAIPCS}
\end{figure*}

It is worth noting that the distribution of profiles in the model C-2DW-100 appears different with respect to the others. This is because in this model the circumbinary disk remains the dominant contributor to the composite broad emission-line profile, while the emission from the mini-disks is suppressed by the optically thick disk wind, characterized by $\tau_0 = 100$. This model therefore includes the profiles associated with the spatially extended circumbinary disk, which are mostly single-peaked, relatively symmetric and narrow, compared to the unattenuated profiles from the mini-disks.

The most notable difference among different models is that the AIP-PS distribution of profiles becomes narrower with the increasing optical depth in the disk wind. Specifically, the measured range of peak shifts calculated in the C-3DW model for SBHBs on circular orbits is between $-8000$ and  $8000\;{\rm km\, s^{-1}}$ in the scenario $\tau_0 = 0$ but is in the narrower range from $-6000$ to $6000\;{\rm km\, s^{-1}}$ in the $\tau_0 = 100$ scenario. Similarly, the AIP-PS distribution of profiles in the E-3DW model becomes narrower with increasing optical depth. 

Figure~\ref{fig:AIPPSb} illustrates how individual profiles change when the optical depth in the disk wind increases. The central panel shows the footprint of the AIP-PS distributions for C-NW and C-3DW-100 models. The gray color represents the distribution in $\tau_0 = 0$ case, while the distribution overlaid in pink traces $\tau_0 = 100$ scenario. In the central panel, the markers ``a" and ``b" trace the location of profiles calculated for the same SBHB configurations with zero and high optical depth, respectively. In all the cases shown, the emission line profiles from the model with $\tau_0 = 100$ tend to concentrate toward the center of the AIP-PS distribution relative to the no wind scenario. It follows that higher optical depth in the disk wind gives rise to more symmetric profiles with a smaller number of peaks. 

The surrounding panels in Figure~\ref{fig:AIPPSb} illustrate the appearance of individual profile pairs. Each composite profile (represented by the black line) is a sum of components contributed by the primary (red), secondary (blue), and circumbinary disk (green). Furthermore, Table~\ref{table:profiles} lists the relevant physical parameters used in the calculation of profiles in Figure~\ref{fig:AIPPSb}. It can be seen that the increase in optical depth transforms the double-peaked profile in panel 1a to a single-peaked profile in panel 1b, mainly because the dominant component contributed by the primary mini-disk becomes single-peaked. A complex profile in panel 4a, which includes comparable contributions from the primary and secondary mini-disks, is reduced to a smoother single-peaked profile in panel 4b. Similarly, a triple-peaked profile in panel 6a, in which the emission from the secondary mini-disk dominates, is reduced to an asymmetric, double-peaked profile in panel 6b. We also list in Table~\ref{table:profiles} the ratio between the maximum flux value for attenuated profiles in panels ``b" ($\tau_0=100$) and the corresponding profiles in panels ``a" ($\tau_0=0$), before they were normalized to 1. The ratio between the two cases is of the order of $10^{-3}$, indicating strong attenuation of the absorbed profiles, relative to the NW scenario. As noted in Section~\ref{sec:NOP}, if profile shapes of SBHBs follow a similar trend as regular AGNs (in terms of the frequency of the single- and double-peaked profiles), then their accretion disks must have outflows with optical depths $\tau_0>1$. In this case, we expect that finding an SBHB characterized by a moderately high value of $\tau_0$ should not be uncommon. For objects with higher than average value of $\tau_0$ however, it may be challenging to disentangle the severely attenuated emission-line profile from the continuum.

As previously mentioned, each profile associated with one of the disks in the triple disk system is subject to Doppler boosting and attenuation due to absorption in the accretion disk wind. The imprints of these two phenomena include boosting of the blue shoulder of an individual profile (Doppler effect) and merging of the peaks of an initially double-peaked profile into a narrower single peak (absorption). These effects can still be recognized in the composite profiles, albeit not as easily, because a combination of three different profiles results in diverse profile shapes. As a consequence, it is not obvious which peak will dominate (or appear weakened) for a given binary configuration until the profile summation is done.

In terms of the relative contribution to the profile flux from individual disks, the mini-disk of the larger SMBH tends to dominate, because it has a larger surface area. An exception to this is scenarios illustrated by the composite profile number 6a/6b, in which the flux from the secondary mini-disk dominates even though $q=1$. This ``additional" line flux arises because of the illumination of the misaligned secondary mini-disk by both AGNs. The contribution to the profile flux from the circumbinary disk is negligible in all configurations, because it is further away from the two AGNs than the mini-disks, and because it is co-planar with the binary orbit, and so it intercepts only a small fraction of the AGN radiation.

%3.3==============================================================================%
\subsection{Dependence of profiles on the physical parameters of the binary}\label{sec:dependence}

In this section we examine whether the profiles produced by the second-generation model preserve distinct statistical properties as a function of the SBHB parameters, as found for the first-generation models. This question is of importance because modification of the broad emission-line profiles by the accretion disk wind may limit their diagnostic power by ``erasing" the imprints of the underlying SBHB configurations. 

As mentioned in \S\,\ref{sec:char}, we use statistical functions (AIP, PS, FWHM, etc.) to construct a multi-dimensional parameter space in which we place the modeled emission-line profiles. Figures~\ref{fig:catAIPPS}, \ref{fig:catFWQMCS}, and \ref{fig:catAIPCS} show the two-dimensional maps, that represent different projections through this parameter space. All maps in these three figures are computed for model E-3DW-1 and are equivalent to Figures~7, 11, and 15 in Paper~I, which contain corresponding maps for the NW model. The distributions from the NW model are shown by the black dashed contours in the top left panel of Figures ~\ref{fig:catAIPPS}, \ref{fig:catFWQMCS}, and \ref{fig:catAIPCS}, for comparison. The remaining panels in these figures show how the properties of modeled profiles vary as a function of the SBHB parameters, such as the alignment of the triple disk system, binary mass ratio, orbital separation, and inclination of the binary orbit relative to the observer. 

Figure~\ref{fig:catAIPPS}, for example, illustrates that profiles from SBHBs with wide orbital separations ($a = 10^6\,M$) tend to be very symmetric and concentrated in the center of the AIP-PS parameter space, while close binaries ($a = 5000\,M$) lead to profiles with a much wider base. Similarly, any SBHB configurations where the mini-disks are coplanar with the binary orbit (and the circumbinary disk, by assumption) are characterized by symmetric profiles with ${\rm AIP}\approx0$, majority of which have the dominant peak shifted toward the blue part of the spectrum, as a consequence of Doppler boosting. The misaligned systems are equally likely to be blue-leaning as well as red-leaning and reside in the range of $-0.4 \lesssim {\rm AIP} \lesssim 0.4$. In contrast, profiles associated with SBHB systems with different mass ratios ($q = 1/10$ and 1) and different orientations of the binary orbit relative to the observer's line of sight ($i = 5^\circ$ and $105^\circ$) show significant overlap in their distributions. Similarly to the NW model, these plots indicate that the most important SBHB parameters that determine the degree of asymmetry and the position of the dominant peak in the emission-line profile are the intrinsic alignment of the triple disk system and the orbital semimajor axis. The only notable difference between the profile distributions in models E-3DW-1 and NW is that in the former, the systems with low $q$ and $i$ show less asymmetry ($-0.3 \lesssim {\rm AIP} \lesssim 0.3$) relative to the NW model ($-0.4 \lesssim {\rm AIP} \lesssim 0.4$). 

Figure~\ref{fig:catFWQMCS} shows FWQM-CS maps calculated for emission-line profiles in the E-3DW-1 model. As in the NW model the profiles exhibit a wide range of centroid shifts, $|{\rm CS}| < 4000\,{\rm km\,s^{-1}}$, and can have broad bases with ${\rm FWQM} < 30,000\,{\rm km \,s^{-1}}$. In both models we find that the location of the centroid is a strong function of $a$ in the sense that profiles from close binaries ($a = 5000\,M$) can have a significantly wider range of CS values relative to the wide binaries ($a = 10^6\,M$). In the E-3DW-1 model the semi-major axis also seems to be the parameter that strongly affects the profile width, since for wide separation binaries ${\rm FWQM} < 17,000\,{\rm km\, s^{-1}}$, considerably lower than for the entire sample of profiles. Indeed, we find that on average the profiles in model E-3DW-1 tend to be narrower than their counterparts in the NW model, as a consequence of modification by the disk wind of finite optical depth.

Figure~\ref{fig:catAIPCS} shows the AIP-CS projection of the parameter space and the profile shapes in the E-3DW-1 model. Similar to the previous maps, the statistical distributions are a strong function of $a$, followed by the degree of triple disk alignment marked by the angles $\theta_1$ and $\theta_2$. This implies that AIP-CS maps calculated based on the second-generation model can still be used as a relatively sensitive diagnostic for these properties. The statistical distributions as a function of $q$ and $i$ are also distinct, so the AIP-CS combination may also be used to constrain these parameters, albeit with a somewhat larger degree of degeneracy. The maps in the E-3DW-1 model therefore preserve the key features seen in the NW model.

In summary, we find that radiative transfer in the disk wind does affect the overall shape of emission-line profiles by making them narrower on average and more symmetric in SBHB systems characterized by low $q$ and $i$.  Despite these differences, all correlations between profile distributions and the SBHB parameters identified in Paper~I are preserved, indicating that their diagnostic power is not diminished. As before, we find that the shapes of modeled emission-line profiles are a sensitive function of the binary orbital separation and the degree of alignment in the triple disk system. The synthetic profiles tend to be less sensitive (or more degenerate with respect) to the SBHB mass ratio and orbital inclination relative to the observer. We furthermore find a large degree of overlap between the models of SBHBs on circular and eccentric orbits and therefore do not expect that the profile shapes alone can be employed as a useful diagnostic of eccentricity. These findings can guide expectations when it comes to the analysis of the spectroscopic SBHB candidates in terms of the diagnostic value and limitations of the emission-line profiles.

%4==============================================================================%
\section{Implications for observational searches}\label{sec:implications}

%4.1==============================================================================%

%%%%%%%%%%%%%%%%%%%%%%%%%%%%%%%%%%%%%%%%%
%%%  FIGURE 8
%%%%%%%%%%%%%%%%%%%%%%%%%%%%%%%%%%%%%%%%%
\begin{figure*}[t]
\centering
\includegraphics[width=0.74\textwidth, clip=true]{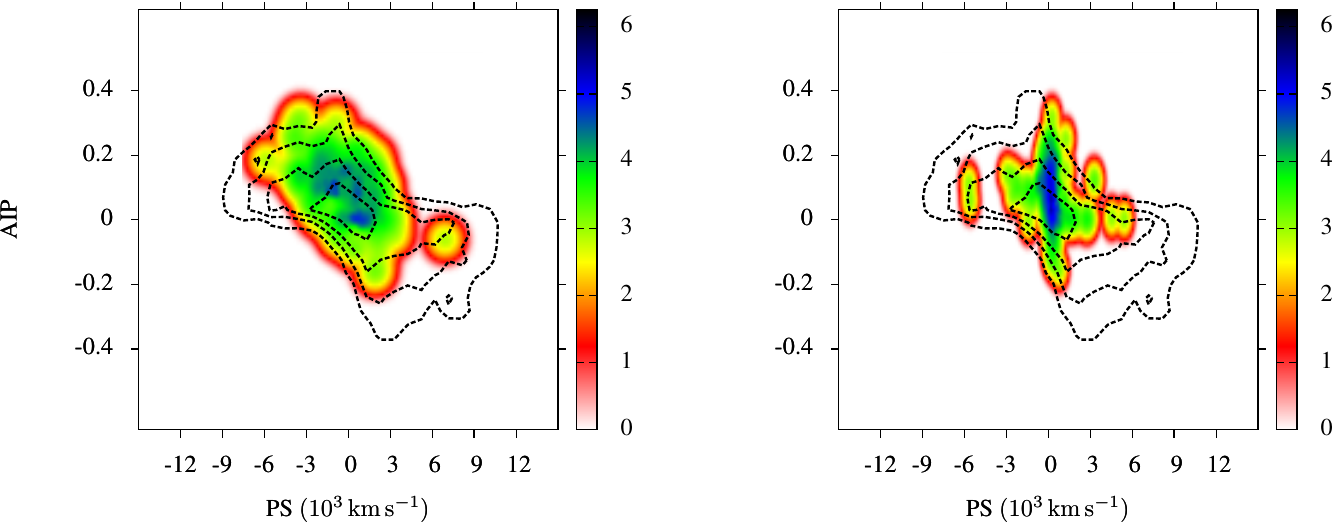}
\caption{AIP-PS maps for profiles associated with the observed SBHB candidates (left) and a control sample of matching AGNs (right) from the E12 search. The maps were adaptively smoothed, normalized to the same profile number and dynamic range as in Figure~\ref{fig:catAIPPS}, in order to facilitate direct visual comparison. Color bar indicates the normalized density of profiles plotted on the $\log$ scale. Black dashed contours correspond to the AIP-CS map in Figure~\ref{fig:catAIPPS} and are drawn in increments of one, from 0.5 (outermost) to 3.5 (innermost).}
\label{fig:obsAIPPS}
\end{figure*}
%
%%%%%%%%%%%%%%%%%%%%%%%%%%%%%%%%%%%%%%%%%
%%%  FIGURE 9
%%%%%%%%%%%%%%%%%%%%%%%%%%%%%%%%%%%%%%%%%
\begin{figure*}[t]
\centering
\includegraphics[width=0.74\textwidth, clip=true]{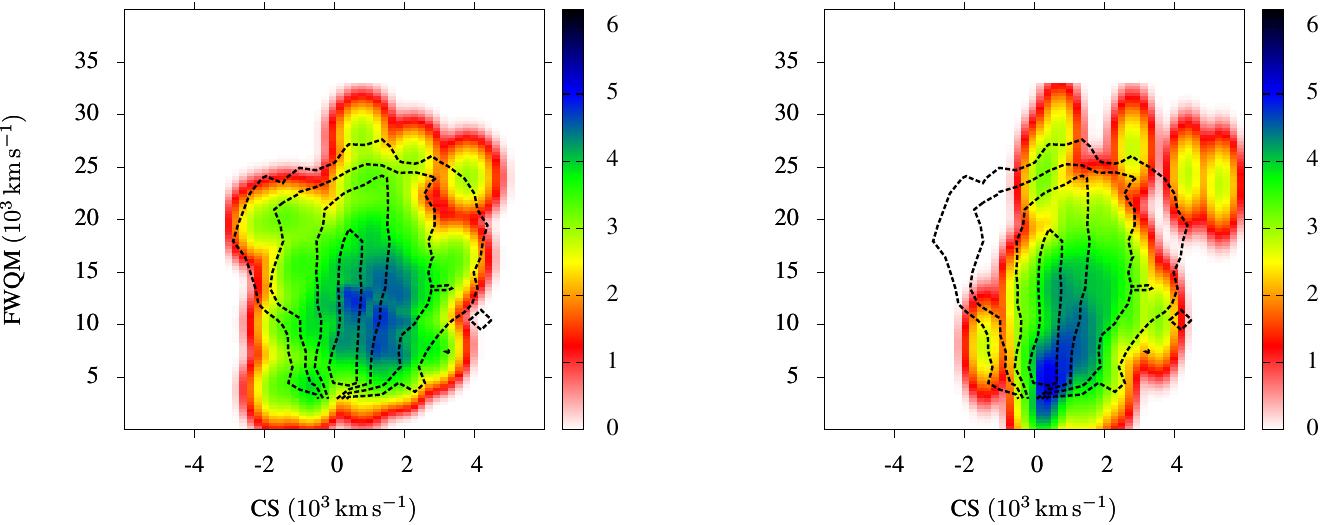}
\caption{FWQM-CS maps for profiles associated with the observed SBHB candidates (left) and a control sample of matching AGNs (right) from the E12 search. Color bar indicates the normalized density of profiles plotted on the $\log$ scale. Black dashed contours correspond to the FWQM-CS map in Figure~\ref{fig:catFWQMCS} and are drawn in increments of one, from 0.5 (outermost) to 3.5 (innermost).}
\label{fig:obsFWQMCS}
\end{figure*}
%
%%%%%%%%%%%%%%%%%%%%%%%%%%%%%%%%%%%%%%%%%
%%%  FIGURE 10
%%%%%%%%%%%%%%%%%%%%%%%%%%%%%%%%%%%%%%%%%
\begin{figure*}[t]
\centering
\includegraphics[width=0.74\textwidth, clip=true]{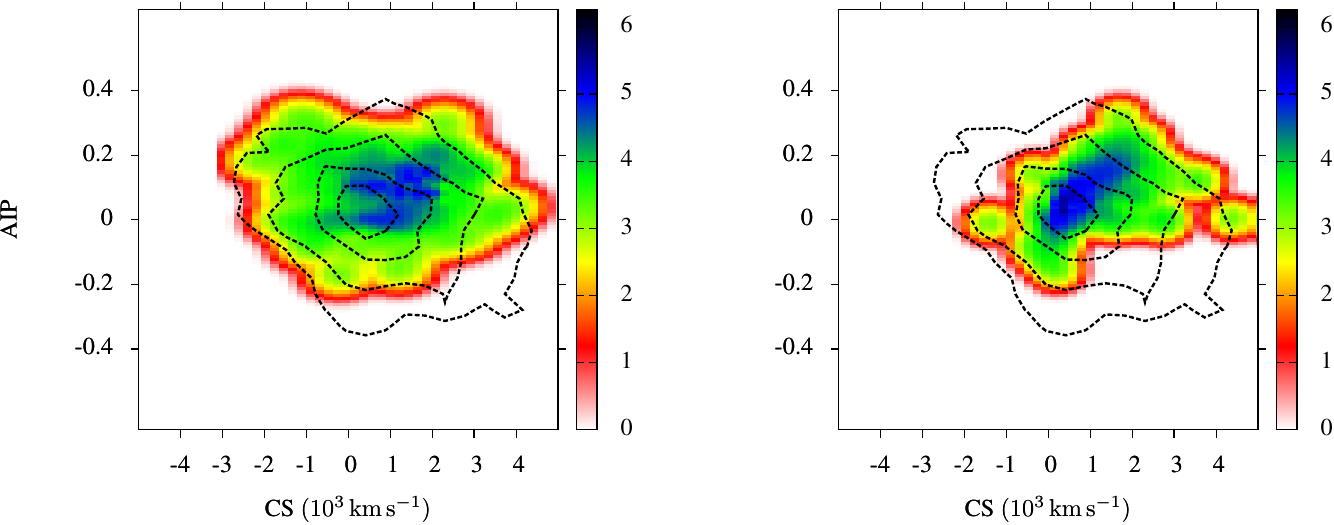}
\caption{AIP-CS map maps for profiles associated with the observed SBHB candidates (left) and a control sample of matching AGNs (right) from the E12 search. Color bar indicates the normalized density of profiles plotted on the $\log$ scale. Black dashed contours correspond to the AIP-CS map in Figure~\ref{fig:catAIPCS} and are drawn in increments of one, from 0.5 (outermost) to 3.5 (innermost).}
\label{fig:obsAIPCS}
\end{figure*}

\subsection{Comparison with SBHB candidates}\label{sec:comparison}

In this section we compare the database of modeled profiles to the emission-lines observed and published as a part of the E12 search for SBHBs. The E12 campaign searched for $z < 0.7$ SDSS\footnote{Sloan Digital Sky Survey} AGNs with broad H$\beta$ lines offset from the rest frame of the host galaxy by $\gtrsim 1000\,{\rm km\,s^{-1}}$.  Based on this criterion, E12 selected 88 quasars for observational followup from an initial group of about 16,000 objects. The followup observations span a temporal baseline from few weeks to 12 years in the observer's frame. Their goal is to measure the epoch-to-epoch modulation in the velocity offset of the H$\beta$ profiles and to test the binarity hypothesis by ruling out any sources in which this modulation is not consistent with the SBHB orbital motion.  After multiple epochs of followup, statistically significant changes in the velocity offset have been measured in 29/88 candidates and reported in the publications mentioned above. At present time, this approach has highlighted several promising cases for further followup but has not yet allowed to rule out the SBHB hypothesis for any candidates.

We use a data set of broad optical emission-lines (drawn from the E12 data set), which at the time of this analysis included 330 multi-epoch spectra of 88 SBHB candidates and 527 spectra for a control sample of 212 matching regular (non-binary) AGNs with similar redshifts and luminosities \citep[see][for description of the candidate and control sample]{runnoe17}. Figures~\ref{fig:obsAIPPS}, \ref{fig:obsFWQMCS}, and \ref{fig:obsAIPCS} show the distribution of these profiles in the AIP-PS,  FWQM-CS, and AIP-CS maps respectively. Because the observed samples of SBHB candidates and regular AGNs contain only a few hundred profiles each, we perform adaptive smoothing of the maps, in order to present them in the the same form as the synthetic data (i.e., as a continuous distribution) that facilitates direct visual comparison. Specifically, the smoothing has been carried out using a two-dimensional (elliptical) Gaussian function with the width scaled linearly with the profile density from 1/10 to 1/3 of the standard deviation of the relevant parameter. These maps can be compared to the corresponding maps for modeled emission line profiles in Figures~\ref{fig:catAIPPS}, \ref{fig:catFWQMCS}, and \ref{fig:catAIPCS}. For easier visual comparison, we also overplot the contours representing these distributions in Figures~\ref{fig:obsAIPPS}, \ref{fig:obsFWQMCS}, and \ref{fig:obsAIPCS}.

For example, inspection of the AIP-PS maps for the observed SBHB candidates and the sample of modeled profiles suggests that both exhibit a negative linear correlation in this projection of the parameter space (i.e., the blue leaning profiles have blueshifted peaks and vice versa). It also follows that, if the observed sample consists of genuine binaries, it must include smaller separation SBHBs ($a \ll 10^6\,r_g$) with misaligned disks. The control sample of AGNs, shown in the right panel of Figure~\ref{fig:obsAIPPS}, on the other hand, does not show such a correlation between the profile asymmetry and peak location and is characterized by profiles with peaks that are predominantly centered on the rest frame of the host galaxy.

%%%%%%%%%%%%%%%%%%%%%%%%%%%%%%%%%%%%%%%%%
%%%  FIGURE 11
%%%%%%%%%%%%%%%%%%%%%%%%%%%%%%%%%%%%%%%%%
\begin{figure*}[t]
\centering
\includegraphics[width=0.7\textwidth, clip=true]{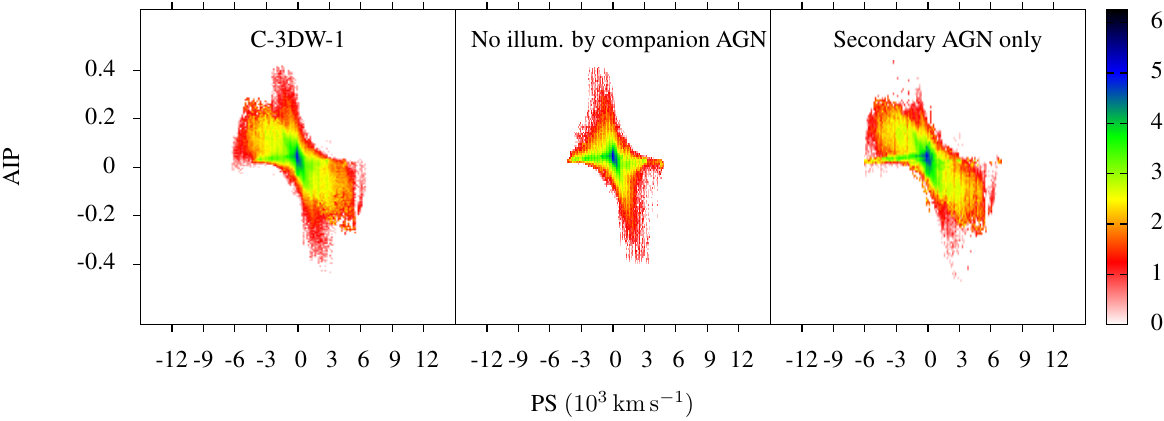}
\caption{AIP-CS maps for the emission-line profiles in the C-3DW-1 model ({\it left}), the scenario in which each AGN is only allowed to illuminate its own mini-disk but not that of the companion SBH ({\it middle}), and in the scenario in which only the AGN associated with the secondary SBH illuminates all three disks, and the primary AGN is assigned zero luminosity ({\it right}). Color bar indicates the density of profiles plotted on a $\log$ scale.}
\label{fig:x1}
\end{figure*}
%
%%%%%%%%%%%%%%%%%%%%%%%%%%%%%%%%%%%%%%%%%
%%%  FIGURE 12
%%%%%%%%%%%%%%%%%%%%%%%%%%%%%%%%%%%%%%%%%
\begin{figure*}[t]
\centering
\includegraphics[width=0.7\textwidth, clip=true]{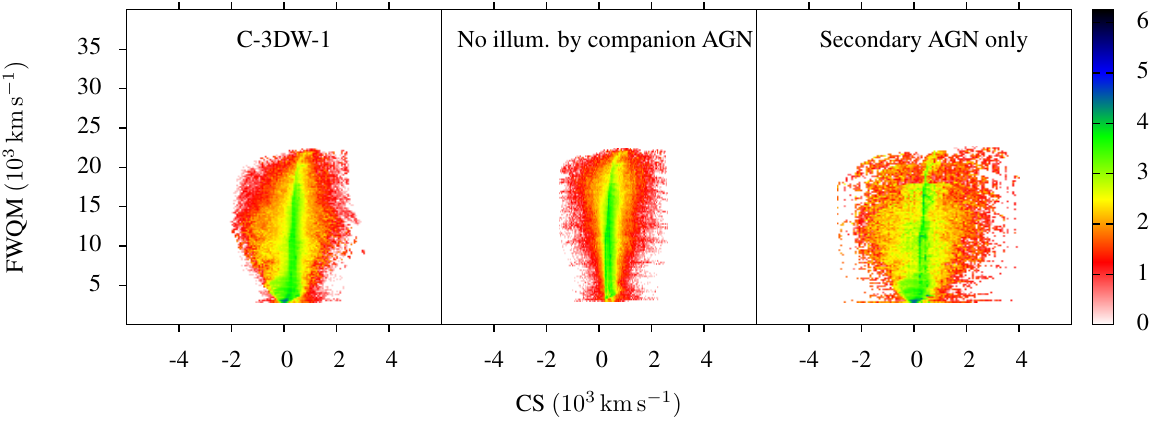}
\caption{FWQM-CS maps with the same legend as in Figure~\ref{fig:x1}.}
\label{fig:x2}
\end{figure*}
%
%%%%%%%%%%%%%%%%%%%%%%%%%%%%%%%%%%%%%%%%%
%%%  FIGURE 13
%%%%%%%%%%%%%%%%%%%%%%%%%%%%%%%%%%%%%%%%%
\begin{figure*}[t]
\centering
\includegraphics[width=0.7\textwidth, clip=true]{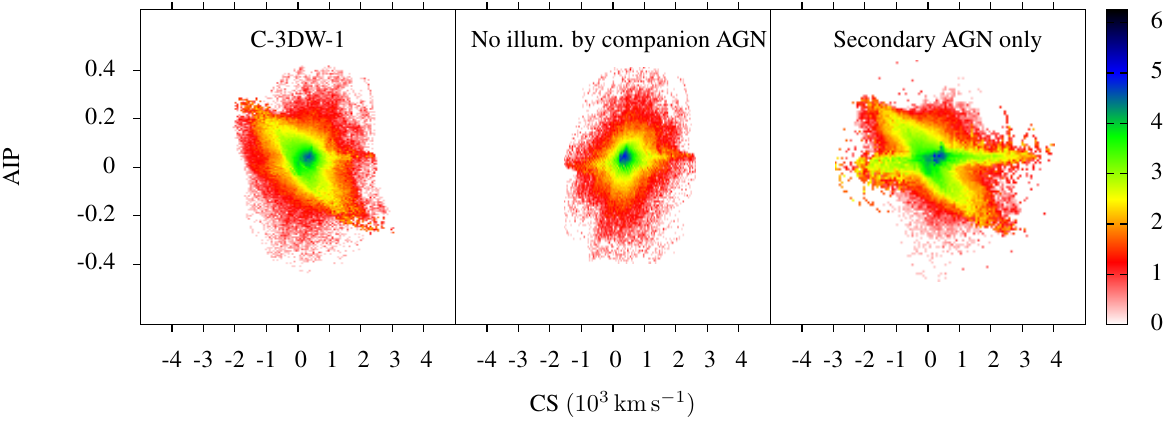}
\caption{AIP-CS maps with the same legend as in Figure~\ref{fig:x1}.}
\label{fig:x3}
\end{figure*}

A comparison of the FWQM-CS maps in Figures~\ref{fig:catFWQMCS} and \ref{fig:obsFWQMCS} reveals that the observed SBHB candidates and the synthetic sample both contain a large fraction of profiles with a relatively broad base, characterized by FWQM$\geq 5000\,{\rm km\,s^{-1}}$. At the same time, most of the profiles in the control sample of AGNs reside at much lower values of FWQM. Figure \ref{fig:obsFWQMCS} also shows that a majority of the SBHB candidates and regular AGNs posses positive CS values, representing profiles with centroids redshifted relative to the frame of the host galaxy by $\sim 1000\,{\rm km\,s^{-1}}$, on average. This prevalence of centroid redshift is also present in the synthetic dataset but the average centroid shift is somewhat smaller for modeled profiles. 

Figure~\ref{fig:obsAIPCS} shows that a majority of the SBHB candidates and regular AGNs posses emission-lines with positive CS and AIP values, indicating preferentially blue leaning, asymmetric profiles with redshifted centroids. In the case of the candidate SBHBs only, the blue leaning profiles with redshifted centroids also tend to have blueshifted peaks (see Figure~\ref{fig:obsAIPPS}), and they account for about $30\%$ of the synthetic database. In general, we find this combination of properties in profiles in which relativistic Doppler boosting plays a role.  A visual comparison of the AIP-CS maps for the observed SBHB candidates and the sample of modeled profiles suggests that, if the observed sample comprises real binaries, it must include compact SBHBs ($a \ll 10^6\,r_g$) with misaligned disks and high mass ratios, as well as the systems with high orbital inclinations relative to the observer's line of sight.

%4.2==============================================================================%

\subsection{Importance of illumination by two active black holes (AGNs)}\label{sec:dualAGN}

All emission-line profiles presented in this work have been calculated assuming that both accreting SBHs can shine as AGNs and illuminate their own mini-disk as well as the two other disks in the system (see Section~\ref{sec:bmodel}). The effect of illumination of one mini-disk by a companion AGN is most pronounced in binaries when their mini-disks are misaligned with the SBHB orbital plane. This geometry allows the companion AGN to effectively illuminate the mini-disk of its neighbor at relatively large incidence angles ($\gtrsim 30^\circ$). As a consequence, in some configurations, the incident flux from the companion AGN on the mini-disk can be several times higher than that of its resident AGN.

As pointed out in Paper I, illumination of the triple-disk BLR by two AGNs can give rise to very asymmetric profiles, with significant peak or centroid velocity shifts. These characteristics are preserved in the second-generation model presented here, and after the emission-line photons are propagated through the accretion disk wind. As a result, all statistical distributions shown in this work and in Paper~I are sensitive functions of the SBHB orbital separation and disk alignment. Because they require a certain degree of geometric misalignment of the mini-disks, the effects of illumination by two AGNs are less ubiquitous (i.e., they affect the shapes of a smaller number of profiles in our database) than the effects on the emission-line photons from the accretion disk wind. When present however, illumination by two AGNs tends to modify the profile shapes more dramatically than the line-driven winds.

This point is illustrated in Figures~\ref{fig:x1}, \ref{fig:x2} and \ref{fig:x3}, which compare the AIP-PS, FWQM-CS and AIP-CS statistical distributions for three different illumination scenarios, respectively. The scenario involving illumination by two AGNs is represented by the C-3DW-1 model, shown in the first panel of all three figures. The second panel illustrates the model in which each AGN is only allowed to illuminate its own mini-disk (but not that of the companion SBH) and both AGNs illuminate the circumbinary disk. The last panel shows the distribution of profiles in the scenario in which only the AGN associated with the secondary SBH illuminates all three disks, and the primary AGN is assigned zero luminosity. The last scenario is of interest for E12 and other spectroscopic searches for binaries, which adopt in their interpretation of the data an assumption that the AGN associated with the secondary SBH is more luminous and outshines the primary.

%%%%%%%%%%%%%%%%%%%%%%%%%%%%%%%%%%%%%%%%%
%%%  FIGURE 14
%%%%%%%%%%%%%%%%%%%%%%%%%%%%%%%%%%%%%%%%%
\begin{figure*}[t]
\centering
\includegraphics[trim=0 0 -20 0, clip, width=0.382\textwidth]{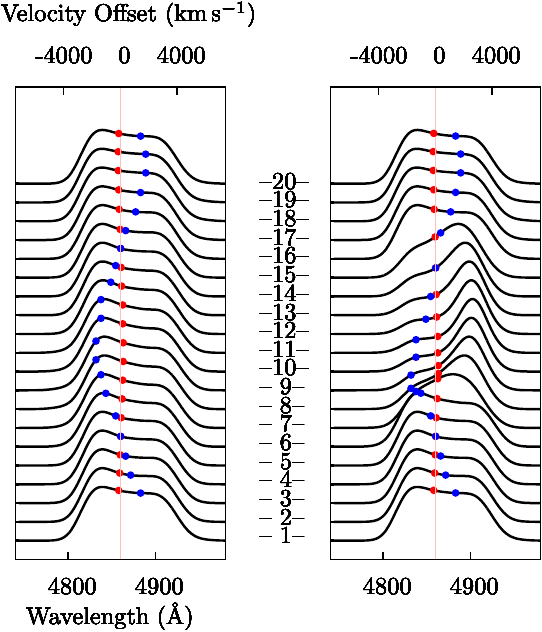} % lbrt
\includegraphics[trim=-20 0 0 0, clip, width=0.58\textwidth]{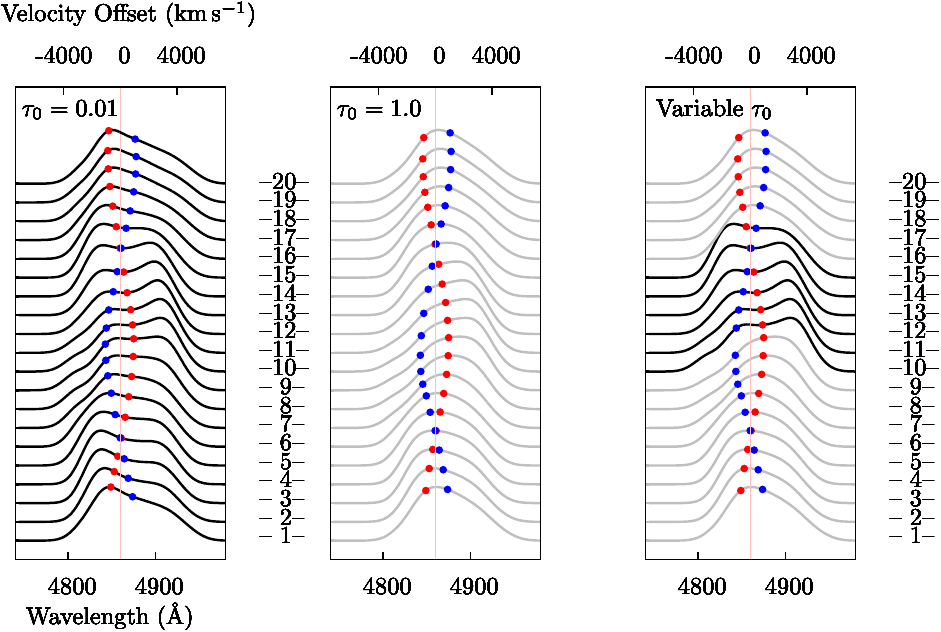} % lbrt
\caption{Temporal evolution of profile shapes over one orbital cycle for an SBHB with $q=1/10$ (panels 1 and 2 from the left) and $q=9/11$ (panels 3, 4 and 5). Red (blue) dots trace the velocity curve of the primary (secondary) SBH. {\it Panel 1:} Profiles calculated for $\tau_0=0.01$ under the assumption that each AGN can only illuminate its own mini-disk and both AGNs illuminate the circumbinary disk. {\it Panel 2:} Profiles calculated for $\tau_0=0.01$ under the (default) assumption that all three disks are illuminated by two AGNs. {\it Panel 3:} Profiles calculated for $\tau_0=0.01$.  {\it Panel 4:} Profiles calculated for $\tau_0=1.0$. {\it Panel 5:} Profiles calculated for optical depth $\tau_0=0.01$ for orbital phases 10--15 and for $\tau_0=1.0$ in all other phases. Other parameters shared by all profiles in this figure are: $a=2500\, M$, $e=0.0$, $R_{\rm in1}=500 \, M_1$, $R_{in2}=500 \, M_2$, $i=20^{\circ}$, $\phi=50^{\circ}$, $\theta_1=15^{\circ}$, $\theta_2=160^{\circ}$, $\phi_1=0^{\circ}$, $\phi_2=180^{\circ}$. The phases 1--20 are equally spaced and phase 1 (20) corresponds to  $f=0^{\circ}$ ($342^{\circ}$). }
\label{fig:fit2}
\end{figure*}

Figure~\ref{fig:x1} illustrates that illumination by a pair of AGNs gives rise to a range of asymmetric profiles with offset peaks which does not happen when only one AGN illuminates each BLR. The latter scenario produces an AIP-PS profile map that resembles the control sample of typical AGNs, shown in Figure~\ref{fig:obsAIPPS}. The scenario with only the secondary AGN as the source of illumination, on the other hand, provides a distribution similar to the C-3DW-1 model and the SBHB candidates. 

In Figure~\ref{fig:x2}, the first and second panels show profile distributions that correspond more closely to that of the control sample of non-binary AGNs in Figure~\ref{fig:obsFWQMCS} than the candidate SBHBs. The ``secondary AGN only" scenario, however, produces a distribution with a much wider range of centroid shifts which does not have an obvious visual resemblance with the profile distributions of either of the observed samples.

In Figure~\ref{fig:x3} the scenarios with illumination from one or both AGNs produce profile distributions comparable to those of the observed sample of SBHB candidates in Figure~\ref{fig:obsAIPCS}. In contrast, the scenario involving illumination only by the secondary AGN produces a distinct distribution consisting of two dominant branches, associated with the SBHB configurations with nearly aligned disks (horizontal branch) and all other misaligned configurations (diagonal branch). The last panel illustrates that the distribution of profiles associated with the SBHB candidates, whether they are true binaries or not, cannot seemingly be reproduced by scenarios in which a single AGN illuminates one or two BLRs, and instead requires more complex emission geometries.

In summary, a simple visual comparison of the statistical distributions of the modeled profiles with the observed samples of SBHB candidates and non-binary AGNs does not provide a clear indication as to which illumination scenario (if any) provides the best description of the data. We discuss the implications of this outcome and describe a simple approach that can be used to compare the simulated profiles to the data in Section~\ref{sec:compare}. It also follows that the effect of illumination by two AGNs can dramatically alter the shapes of emission-line profiles for a subset of SBHB configurations in which the multi-disk BLR is not co-planar. The non-axisymmetric illumination pattern by two AGNs implies that the emissivity distribution of their BLRs is a function of the binary orbital phase. Such emissivity distribution gives rise to emission-line profiles with shapes that can vary on time scales shorter than the SBHB orbital period. We describe the implications of the varying illumination and obscuration for the profile variability in the next section.

%4.3==============================================================================%

\subsection{Temporal variability of the modeled line profiles}\label{sec:tevolution}

In our model temporal variability of the emission-lines can arise on the orbital time scale when profile modulation is associated with the orbital motion of the SBH mini-disks and/or with the changing illumination pattern by two AGNs. Alternatively, profile variability can arise on time scales different from the orbital time scale if it is associated with the change in optical depth of the wind along the line of sight. While we do not explicitly model different temporal sequences for profiles with varying $\tau_0$, they can be created by choosing the appropriate SBHB and wind optical depth configurations from the synthetic database.

The first panel of Figure~\ref{fig:fit2} shows 20 orbital phases within one orbital cycle of an SBHB system with $q=0.1$ and remaining parameters as shown in the caption. Red dots trace the modulation of the emission line profile associated with the orbital motion of the primary SBH and the blue dots trace the secondary. An important assumption made in the calculation of these profiles is that each AGN can only illuminate its own mini-disk and both AGNs illuminate the circumbinary disk,  a scenario presented in \S\,\ref{sec:dualAGN}. In this case, there are no significant changes to the shape or width of the composite profile and the wavelength shift is relatively small and invisible to the eye. This can be understood because at $q=0.1$ the emission from the mini-disk of the primary SBH dominates over the mini-disk of the secondary, since the ratio of their fluxes scales as $F_2/F_1 \propto q^2$ (see equation 38 in Paper~I). Note that the primary mini-disk dominates even if one accounts for the inversion of accretion rates in unequal mass binaries, where accretion is expected to occur preferentially onto the smaller of the SBHs. Specifically, in the $q=0.1$ configuration considered above, the accretion rate onto the secondary SBH is $\sim 5$ times higher than that of the primary\footnote{See equation 3 in Paper~I for the description of the SBHB accretion rate ratio.}. Therefore, even though the more luminous secondary AGN in our model boosts the flux ratio by a factor of $\sim 5$, this is still insufficient for the secondary BLR to outshine the primary BLR in the Balmer lines. At the same time, the contribution to the profile flux from the circumbinary disk is negligible in a majority of configurations in our model, as illustrated in Figure~\ref{fig:AIPPSb}.

The second panel of Figure~\ref{fig:fit2} shows emission-line profiles associated with the same SBHB configuration but in this case, illumination of all three disks by both AGNs is allowed. Note that this is a default assumption used in the calculation of all profiles in our database of modeled profiles. Because this system consists of misaligned mini-disks (as indicated by the angles $\theta_1$ and $\theta_2$), the geometry of the system allows for illumination of the primary BLR by the secondary AGN. This effect leads to a significant change in the profile shape over the limited portion of the orbital cycle, in phases 7--15. Therefore, even though the contribution to the line flux from the secondary BLR is negligible, the illumination by the secondary AGN is not. As a result, the spectroscopic signatures of SBHBs in circumbinary disks might be unique even when their mass ratio is very low. 

The remaining three panels of Figure~\ref{fig:fit2} show different phases of the same SBHB configuration with $q=9/11$. The effect of illumination by both AGNs is accounted for in all profiles shown in these three panels. This system is representative of SBHBs with $q\sim 1$, surrounded by mini-disks that make comparable contributions to the composite emission-line profile. Because of their individual modulation in wavelength over time (indicated by the red and blue dots), the composite profile varies in both shape and width throughout the orbital cycle, but exhibits insignificant shift with respect to the rest frame of the host galaxy. To illustrate the effect of the optical depth on the profile shapes in this scenario we show the emission-lines calculated for $\tau_0=0.01$ and $1.0$ in the third and fourth panels, respectively. The primary effect of the increased optical depth in the disk wind is a transformation from a broader and occasionally double-peaked emission-line profile into a smoother and narrower single-peaked profile. 

To emulate a variable optical depth, in the fifth panel we show the profiles calculated for $\tau_0=0.01$ in phases 10--15 and the remaining profiles corresponding to the higher optical depth. Note that similar changes have been seen in the broad Balmer emission lines of some AGNs that are not SBHB candidates, which have been observed to fluctuate between a double-peaked and a single-peaked profile: NGC 5548 \citep{peterson99,shapovalova04,sergeev07}, Pictor A \citep{halpern94,sulentic95}, and Ark 120 \citep{alloin88,marziani92}. The disk-wind with changing optical depth has been suggested as a viable explanation for the appearance of these sources \citep{flohic12}.

In summary, we find that orbital modulation in a binary with small $q$, in which both black holes are shining as AGN, results in smaller radial velocity offsets of the emission-line profiles, determined by the velocity curve of the primary. Equal mass binaries exhibit no radial velocity offsets in binary AGN systems due to symmetry but their profiles show plenty of variation in shape on the orbital time scale. The disk wind has a weak impact on the radial velocity offsets or profile variability and its main effect is to make the profiles smoother and single-peaked. The most dramatic variations in shape are caused by illumination of one mini-disk by a companion AGN. The properties of the profile that show the biggest changes are the asymmetry (AIP) and peak shift (PS), whereas FWHM is not so strongly impacted. The effect is only noticeable over a fraction of the orbital cycle, suggesting that if present in real SBHBs, only a fraction of binaries should be affected by it at any given time.

%4.4==============================================================================%

\subsection{Implications for the observed variable profiles}\label{sec:tevolutionobs}

The configuration in the fourth panel of Figure~\ref{fig:fit2} is chosen to mimic the variability of the H$\beta$ profiles observed in NGC~5548 \citep[see Figure 5 of][]{li16}, which has been proposed as a nearby SBHB candidate with an orbital period of 14--16 years \citep[see also][]{bon16}. This orbital period corresponds to a $10^8\,M_\odot$ SBHB with an orbital separation close to $2500\,M$, similar to the example shown here. It is worth noting that this apparent similarity in profile shapes does not provide a proof of binarity for NGC~5548, because it does not rule out other non-SBHB mechanisms, which in principle may produce a similar sequence of profiles. However, should the binary hypothesis for NGC~5548 be confirmed, our model suggests that this is likely to be an SBHB with $q\sim 1$ and misaligned mini-disks, which allow for the changing illumination by the two AGNs.

The emission-line profiles that exhibit significant change in their shapes from one epoch of observation to another, similar to those shown for the high mass ratio SBHB in Figure~\ref{fig:fit2}, represent a practical challenge for the spectroscopic searches for SBHBs which seek to measure the wavelength shift of the entire profile. As discussed in publications reporting on the E12 search, evolution of profile shape makes it very difficult to discern the shift of the bulk of the profile, as the former may mimic the latter. For example, \citet{runnoe17} report reliable and statistically significant measurements of radial velocity changes in 29/88 SBHB candidates, which exhibit no variability in the shape of the broad $H\beta$ profile over the length of the monitoring campaign. In this context, E12 hypothesize that these 29 profiles correspond to the SBHB systems, where the mini-disk of the secondary SBH is the dominant contributor to the $H\beta$ line flux\footnote{The SBHB interpretation adopted by E12 and other spectroscopic searches also leaves room for the primary mini-disk to be the dominant contributor to the emission-line, given a binary with the mass larger by a factor $q^{-3}$.} and as a result, the composite profile does not change in shape or width over the observed portion of the orbital cycle. 

In the context of our model the 29 candidates with relatively stable profile shapes may correspond to configurations in which the evolution of the profile shapes is slower and is not discernible on the monitoring time scale of few to ten years. This may indicate a longer period binary (with orbital period of a few hundred years) or a system where either the primary or the secondary BLR dominate the emission-line flux over some fraction of the orbit. Since this behavior is consistent with a relatively diverse set of SBHB configurations in our model we cannot make more detailed inferences about the corresponding orbital separations and mass ratios. 

Furthermore, 49/88 SBHB candidates in the E12 sample are characterized by variable profiles which preclude the radial velocity measurements, and the rest show no measurable radial velocity changes. By implication, the 49 candidates with variable profile shapes are consistent with $q>0.1$ configurations modeled in this work, and with small separation binaries ($a\lesssim 10^4\,M$) with misaligned disks, in which changing illumination by the two AGNs plays an important role over the observed portion of the orbit.

%5==============================================================================%
\section{Discussion}\label{sec:discussion}

Because the model presented here is developed from the same basic principles as the first-generation model presented in Paper~I, many of the assumptions are shared by both. We direct the reader to Paper~I for discussion of the implications of these simplifying assumptions and only address the new aspects of the model here.

One important assumption of the first-generation model, that is relaxed in this work, pertains to the effect of feedback from the binary AGN on the emission signatures of SBHBs in circumbinary disks. Specifically, in the second-generation model we assume that the radiation pressure from the two AGNs is capable of driving winds and outflows, which change the effective line optical depth of the emitting gas. As described in previous sections, this results in simpler, mostly single-peaked emission line profiles, which resemble those of observed SBHB candidates and AGNs in general. However, this addition to the model inevitably makes it more complex, as it requires the introduction of new parameters to describe the properties of the accretion disk wind. The primary effect of the increase in the number of parameters is the increased degeneracy in the relationship between the properties of the emission-line profiles and the underlying SBHB parameters. We show in \S\,\ref{sec:dependence} that the correlations between profile distribution functions and the SBHB parameters identified in the first-generation model are nevertheless preserved, indicating that their diagnostic power is not diminished.

Along similar lines, the calculation of radiative transfer requires several assumptions about the properties of the accretion disk wind. The main one is that the Sobolev approximation is applicable to the accretion disk winds arising in the BLRs of SBHB systems. In this regime the photons that are not absorbed within one Sobolev length from the point of emission (or in the case of a disk, from the emission layer) can escape to infinity, provided that the velocity of the wind projected onto the line of sight is monotonically increasing (see Appendix~\ref{sec:Q}). This condition is likely satisfied in regular AGNs, in which accretion disk winds are expected to accelerate radially out \citep{proga00, proga04}. 

In our model we also make a simplifying assumption that the two mini-disks have the same disk-wind optical depth. It is therefore worth understanding how different the physical conditions can be in the two SBH mini-disks, which tend to contribute most of the flux to the composite H$\beta$ line profile. In our model, which is motivated by hydrodynamic simulations of SBHBs in circumbinary disks, the largest contrast between the SBH mass ratios is $\dot{m} = \dot{M}_2/\dot{M}_1 \approx 5$, for circular binaries with mass ratio $q=1/10$ (see equation~3 in Paper~I). In terms of the Eddington normalized mass ratios this implies $\dot{m}/q \approx 50$. If the emission-lines from SBH mini-disks respond to the continuum radiation in the same way as in regular AGNs, the higher relative luminosity of the secondary AGN would result in a lower equivalent width of the H$\beta$ emission-line profile contributed by the secondary mini-disk \citep[correlation known as the Baldwin effect;][]{baldwin77}. In the context of our calculation, this implies a reduction in the contribution to the composite H$\beta$ profile from the secondary mini-disk, an effect not captured by our model. The effect is weaker for circular binaries with larger SBH mass ratios and for all eccentric SBHBs, and it disappears in all configurations when $q=1$.

The geometry and kinematics of radiation driven outflows in SBHB systems are unknown. In the absence of any other constraints we assume that in a binary the disk wind driven by each AGN extends over the entire surface of its BLR and is not affected by the wind from the companion disk. We also assume that the properties of the disk wind (e.g., mass density and velocity profile) in each BLR are similar to regular AGNs. If, on the other hand, the outflows from the two mini-disks are interacting and colliding, the velocity field in the wind can become non-monotonic and the local escape probability defined in equation~\ref{eq:escape1} would cease to be a good description of radiation transport. One would instead have a more complex distribution of the line optical depth across one or both mini-disks, which would in turn give rise to more complex profiles \citep{rybicki78}.

It is also not clear whether the outflows can extend from the mini-disks into the circumbinary disk, especially in configurations in which they are not co-planar. In this work we consider configurations in which the circumbinary disk is (a) either not affected by the disk wind (2DW models) or (b) has an accretion disk wind which is radial and axisymmetric, as if driven by a single, central AGN (3DW models). When it comes to the 2DW suite of models, a model which is clearly inconsistent with the observed sample of SBHB candidates (and appears more like regular AGNs) is C-2DW-100, in which the emission from the mini-disks is suppressed due to the high optical depth ($\tau_0 = 100$) and the emission from the circumbinary disk dominates. In the case of the 3DW models, the circumbinary disk affected by a wind makes a negligible contribution to the flux of the composite profile and hence, we do not expect the assumptions in (b) to strongly affect our results.

The dependence on the physical parameters of the SBHB makes the broad emission-line profiles a promising diagnostic once a sample of genuine sub-parsec binaries is available. At this point, a comparison of the probability density distributions of modeled profiles with those from observed SBHB candidates (\S~3.3 and 4.1) suggests that, if the observed sample is made up of real binaries, it must include smaller separation SBHBs ($a \ll 10^6\,r_g$) with misaligned disks and high mass ratios. While this is intriguing, a visual comparison of the distributions does not provide conclusive evidence that they overlap or are drawn from some larger, common distribution. We nevertheless perform a simple comparison and find that the shapes of the emission-line profiles from a sample of observed SBHB candidates are more consistent with the binary model than are regular AGNs (see Appendix~\ref{sec:compare} for more details).

A point worth reiterating is that other physical processes can potentially mimic the emission signatures of SBHBs discussed here. These include but are not limited to the recoiling SBHs \citep{blecha16} and local and global instabilities in single SBH accretion disks that can give rise to transient bright spots and spiral arms \citep{flohic08,lewis10}. In that sense, the model described in this paper can be used to interpret observed emission line profiles in the context of the SBHB model but cannot be used to prove that they originate with genuine SBHB systems. For example, profiles of SBHB candidates observed in multiple epochs can be compared against the synthetic database individually, in order to determine the likelihood distribution for underlying SBHB parameters for each profile independently. The entire time series of observed profiles can also be compared against the time series of matching modeled profiles as an added consistency check for the inferred SBHB parameters. We defer this type of analysis to Paper~III of the series.

%6==============================================================================%

\section{Conclusions}\label{sec:conclusions}

We present an improved semi-analytic model to describe the spectral emission-line signatures of sub-parsec SBHBs. This work is a follow-up to Nguyen \& Bogdanovi\'c (2016), in which we showed that the broad optical emission-line profiles associated with SBHBs in circumbinary disks can in principle be used as diagnostic tools to constrain the distribution of the binary semi-major axis, the mass ratio and the alignment of the triple-disk system. 

The second-generation model improves upon the treatment of radiative transfer by taking into account the effect of the radiation driven, accretion disk winds on the modeled emission-line profiles. Under the influence of accretion disk winds the emission-line profiles appear narrower, more symmetric, and predominantly single-peaked. All correlations between the profile shape parameters and SBHB parameters identified in the first-generation model are nevertheless preserved, indicating that their diagnostic power is not diminished.  Analysis of 42.5 million modeled emission-line profiles reveals that their shapes are a better indicator of the binary orbital separation and the degree of misalignment in the triple disk system, and are less sensitive to the SBHB mass ratio and orbital eccentricity. These findings can guide expectations related to the diagnostic value and limitations of the emission-line profiles in spectroscopic searches for SBHBs. 

The emission-line profiles presented in this work have been calculated assuming that each accreting SBH can shine as an AGN and illuminate its own mini-disk as well as the two other disks in the system. The illumination by two AGNs gives rise to the characteristic emission-line profiles, with shapes distinct from those produced by a single BLR illuminated by the central AGN. Moreover, as a consequence of the evolving illumination pattern from the two AGNs, the emission-line profiles associated with the SBHBs in circumbinary disks can exhibit significant variability on the orbital time scale of the system. We identify this as a key spectroscopic signature of the SBHB systems but cannot rule out a possibility that these features are mimicked by transient bright spots and spiral arms in single SBH accretion disks. We defer this question to a separate paper.

We perform a comparison of the modeled emission-line profiles with those for the observed SBHB candidates and a control sample of (non-binary) AGNs from the E12 sample. We find that the profile shapes from a sample of SBHB candidates are more consistent with the binary model than are regular AGNs. Furthermore, the comparison of the modeled profiles with the SBHB candidates indicates that, if the observed sample comprises genuine binaries, it must include systems with smaller separations ($a \ll 10^6\,r_g$), comparable masses, and misaligned (or possibly warped) mini-disks. Similarly, if all candidates are shown to be genuine SBHBs, this would require their BLRs to be enshrouded in disk winds with optical depth $\gtrsim 0.1$.

Finally, we emphasize that the model described in this paper cannot be used to prove that the observed SBHB candidates are real binaries but it can be used to interpret the observed emission-line profiles once a sample of confirmed SBHBs is available. In anticipation of this outcome, we envision a framework within which the profiles of SBHBs obtained in multiple epochs of observation can be mapped into the synthetic database individually and as a time sequence, in order to determine the most likely values of the underlying SBHB parameters. We defer this analysis to Paper~III of the series.

\acknowledgements
We thank the anonymous referee for a careful reading of the manuscript and thoughtful comments that helped to improve this work. This research was supported by the National Aeronautics and Space Administration under Grant No. NNX15AK84G issued through the Astrophysics Theory Program and by the Research Corporation for Science Advancement through a Cottrell Scholar Award. ME, JCR, and SS acknowledge support from grant AST-1211756 from the National Science Foundation and an associated REU supplement. TB acknowledges support from grant AST-1211677 from the National Science foundation during the early stages of this work. Part of this work was performed at the Aspen Center for Physics, which is supported by National Science Foundation grant PHY-1607611. Numerical simulations presented in this paper were performed using the high-performance computing cluster PACE, administered by the Office of Information and Technology at the Georgia Institute of Technology.

%Appendix A==============================================================================%
\appendix
\label{appendix}
\section{A. Implementation of the disk wind Model}\label{sec:DW}

In this section we describe the new elements in the calculation of the broad optical emission-line profiles, introduced as a part of the disk wind model. We refer the reader to Paper~I for the description of geometry of a triple disk system and main steps in the calculation of the composite emission-line profiles. The same geometry and procedure are also used in this work.  As before, we adopt the disk emitter model \citep{chen89, chen89b, eracleous95} to describe the line flux emitted by each individual disk in the system. In this context the flux of the broad emission-line profile measured in the observer's frame is an integral over the surface of the emitting disk defined in terms of the properties in the disk frame:
\begin{equation}
\label{eq:flux1}
F_{\nu}(\nu_{\rm obs})=\frac{M^2\nu_0}{d^2} 
\int\limits_{0}^{2\pi} \int\limits_{\xi_{\rm in}}^{\xi_{\rm out}}   
I_{\nu}(\xi,\varphi,\nu_{\rm turb}) D_{\rm rot}^3 \left( 1-\frac{2}{\xi} \right) ^ {-\frac{1}{2}} \xi\,  d\xi\, d\varphi \;\;.
\end{equation}
Here we assume that the emitting disk is geometrically thin and that emission arises from the $\theta \approx 90^\circ$ plane. $M$ is the mass of the central black hole, $\nu_0$ is the rest frequency of the emission line, $d$ is the distance from the center of the disk to the observer, $\xi = r/M$ is the radius in the disk in dimensionless units, $\xi_{\rm in}$ and $\xi_{\rm out}$ are the inner and outer edges of the emission region, respectively, and $\varphi$ is the azimuthal angle measured in the plane of the disk (note that $\varphi$ is different from  $\phi$ defined in Section~\ref{sec:bmodel} and Table~\ref{table:parameters}). $D_{\rm rot}$ is the relativistic Doppler factor.

$I_{\nu}(\xi,\varphi, \nu_{\rm turb})$ is the specific intensity of light emitted at a polar coordinate ($\xi, \varphi$) and frequency $\nu_{\rm turb}$. In the disk wind model
\begin{equation}
\label{eq:intensity1}
I_{\nu}(\xi,\varphi,\nu_{\rm turb})= \frac{\beta_e}{4\pi} 
\frac{\epsilon(\xi,\varphi)}{(2\pi)^{1/2} \sigma} \, 
\exp[{-(\nu_{\rm turb}-\nu_0)^2/2\sigma^2}] \;\;.
\end{equation}
Recall that the emissivity of a BLR illuminated by a single, central AGN is axisymmetric and thus independent on $\varphi$. In our model however, every disk in the system is illuminated by two AGNs, giving rise to a non-axisymmetric emissivity pattern encoded in $\epsilon(\xi,\varphi)$, as defined in equations 28--31 of Paper~I. The broadening of the emission line profiles is assumed to be due to the turbulent velocity of the disk, $\sigma$, and is described as a Gaussian distribution around $\nu_0$. 

The new element of the calculation is the modification of the disk emissivity by absorption of line photons in the disk wind. The absorption is characterized by a photon escape probability, $\beta_e(\vec{r},\,\vec{\hat{s}})$,  which represents the probability for an emission-line photon, emitted from the surface of the disk at a location $\vec{r}$, to escape the wind in the direction of the line of sight defined by the unit vector $\vec{\hat{s}}$. The escape probability is a function of the line optical depth of the wind (equation~\ref{eq:escape1}), which in turn depends on the density and velocity field of the wind. We describe how are these properties modeled in the next sections.

\subsection{A.1. Velocity field of the disk wind}\label{sec:DWvelocity}

We describe the three-dimensional velocity field of the wind in terms of the spherical velocity components $v_r$, $v_\theta$ and $v_{\varphi}$, defined with respect to the center of the disk, which is located in the $xy$ plane and has the angular momentum vector aligned with the positive $z$-axis. Following the formalism developed for isotropic stellar winds we express the poloidal component of the velocity along a streamline, $\vec{v}_p = \vec{v}_r + \vec{v}_\theta$, as an increasing function of distance from the launching point 
\begin{equation}\label{eq_velr}
v_p(r)=v_{\infty} \left(1- b\, \frac{r_f}{r}\right)^{\gamma}
\end{equation}
\citep[see][]{kudritzki00}. The parameter $b=1 - [v_p(r_f)/v_{\infty}]^{1/\gamma}$ is related to the ratio between the initial velocity of the wind at the launching point (i.e., at the foot-point of the streamline) and its terminal velocity far downstream. For the purposes of this work we choose $b = 0.7$. The parameter $\gamma$ is expected to range from $1.06$ for quasars to $1.3$ for Seyferts \citep[see][]{mc95}, and we choose $\gamma=1.2$ as a representative intermediate value. Following \citet{cm96} and \citet{flohic12}, we also adopt $v_{\infty}(r_f)= 4.7(GM/r_f)^{1/2}$ and the velocity components 
\begin{equation}\label{eq:vtheta}
v_r = v_p\cos\lambda\;\;\, ,  \;\;\; v_{\theta} = - v_p\sin\lambda\,\,\,\,\,  {\rm and}\,\,\,\,\,  v_{\varphi}= \left(\frac{GM}{r}\right)^{1/2} \;\;\; ,
\end{equation}
respectively. Here $\lambda = \lambda(R_{\rm in}) (R_{\rm in}/r)$ is the wind opening angle, measured between the streamline and the disk, and $\lambda(R_{\rm in})= 10^\circ$, as defined in equation~\ref{eq:wangle1} and shown in Figure~\ref{fig:geom}. These choices imply the finite poloidal and azimuthal velocity at the foot-point of the wind, $v_p(r_f) = v_{\varphi}(r_f) = (GM/r_f)^{1/2}$, which combined in quadrature equal the escape velocity from the SBH. Moreover, all velocity components are characterized by non-zero velocity gradients, as required by the Sobolev method. Note also that the symmetry of a single disk system requires that $v_{\theta}=- v_p\sin\lambda$ when $i<90^{\circ}$, and $v_{\theta}= v_p\sin\lambda$ when $i>90^{\circ}$.

It is worth noting that our choices for $b$ and $\gamma$ are different from those previously used in the literature. For example \citet{cm96} and \citet{flohic12} choose $b=1$, which leads to a simplification of the model because $v_p(r_f)$ vanishes in that case, implying that the wind starts with the zero velocity and accelerates outwards. This simplification in turn necessitates that the value of $\gamma$ is exactly 1, because for any other value the gradient at the foot-point of the streamline $\partial v_p/\partial r = 0$, thus violating the requirement for high velocity gradients in the Sobolev method.

\subsection{A.2. Velocity gradient of the wind along the line of sight, Q}\label{sec:Q}

In this section we outline the calculation of the velocity gradient of the wind along the line of sight, given the velocity field of the wind defined above. As laid out in Section~\ref{sec:dwmodel}, the velocity gradient can be expressed as an inner product, $Q=\vec{\hat{s}}\cdot\vec{\Lambda}\cdot\vec{\hat{s}}$, of the strain tensor along a given line of sight. Without loss of generality, we assume that a distant observer is located in the $xz$ plane at an inclination angle, $i$, relative to the $z$-axis. Hence, the direction of the line of sight, pointing from any point on the disk to the distant observer is given by
\begin{equation}
\vec{\hat{s}}=\cos i\,\vec{\hat{z}}+\sin i\,\vec{\hat{x}}=\sin i\cos\varphi\,\vec{\hat{r}}-\cos i\,\vec{\hat{\theta}}-\sin i\sin\varphi\,\vec{\hat{\varphi}} \;\;.
\end{equation}
Here we use the relationships between the unit vectors in the Cartesian and spherical-polar coordinate systems, $\vec{\hat{x}}=\sin\theta\cos\varphi\,\vec{\hat{r}}+\cos\theta\cos\varphi\,\vec{\hat{\theta}}-\sin\varphi\,\vec{\hat{\varphi}}$ and $\vec{\hat{z}}=\cos\theta\,\vec{\hat{r}}-\sin\theta\,\vec{\hat{\theta}}$. Setting $\theta=90^{\circ}$ for any point on the surface of the disk and expanding the inner products yields
\begin{equation}\label{eq:Q1}
Q = \sin^2i \left[ \cos^2 \varphi\,\Lambda_{rr}-2\sin \varphi \cos \varphi\,\Lambda_{r\varphi} + \sin ^2 \varphi \,\Lambda_{\varphi\varphi} \right] - \lvert \cos i \rvert \left[ 2 \sin i\cos \varphi\,\Lambda_{r\theta}- \lvert \cos i \rvert \,\Lambda_{\theta\theta}-2\sin i \sin \varphi\,\Lambda_{\theta\varphi} \right] \,,
\end{equation}
where the absolute value of $\cos i$ ensures the symmetry of solutions above and below the accretion disk, such that $Q(i)=Q(180^{\circ}-i)$. It is worth noting that equation~\ref{eq:Q1} is consistent with the equation~8 in \citet{chajet13}, who pointed out and corrected the sign error affecting the $\cos i$ terms in the expression for $Q$ in \citet{mc97}, in their equation~15.

For an azimuthally symmetric disk wind ($\partial/\partial\varphi = 0$) with a small opening angle ($\cos\lambda \approx 1$), the components of the symmetric strain tensor $\vec{\Lambda}$ in spherical coordinates are:
\begin{equation}
\Lambda_{rr}=
\frac{\partial v_r}{\partial r}=
\frac{C_1}{\kappa\,M}\, \xi^{-3/2} \,\,,
\label{eq_Lrr}
\end{equation}
\begin{equation}
\Lambda_{\theta\theta}=
\frac{1}{r}\frac{\partial v_{\theta}}{\partial \theta} + \frac{v_r}{r}=\frac{\partial v_r}{\partial r} + \frac{v_r}{r}=
\frac{C_1+C_2}{\kappa\,M}\,\xi^{-3/2}\,\,,
\end{equation}
\begin{equation}
\Lambda_{\varphi\varphi}=
\frac{1}{r \sin \theta}\frac{\partial v_{\varphi}}{\partial \varphi}+\frac{v_r}{r}+\frac{v_{\theta} \cot \theta}{r}=\frac{v_r}{r}=\frac{C_2}{\kappa\,M}\,\xi^{-3/2}\,\,,
\end{equation}
\begin{equation}
\Lambda_{\theta\varphi}=
\frac{\sin \theta}{2r}\frac{\partial}{\partial \theta}\left(\frac{v_{\varphi}}{\sin \theta}\right)+\frac{1}{2r\sin \theta}\frac{\partial v_{\theta}}{\partial \varphi}=\frac{1}{2r}\frac{\partial v_{\varphi}}{\partial \theta}=\frac{1}{4\sin \lambda}\frac{v_{\varphi}}{r}=
\frac{1}{4M\, \sin \lambda}\, \xi^{-3/2}\,\,,
\end{equation}
\begin{equation}
\Lambda_{r\varphi}=
\frac{1}{2r \sin \theta} \frac{\partial v_r}{\partial \varphi}+\frac{r}{2}\frac{\partial}{\partial r}\left(\frac{v_{\varphi}}{r}\right)=
\frac{r}{2}\frac{\partial}{\partial r}\left(\frac{v_{\varphi}}{r}\right)=
-\frac{3}{4M}\,\xi^{-3/2}\,\,,
\end{equation}
\begin{eqnarray}
\Lambda_{r\theta}=
\frac{r}{2}\frac{\partial}{\partial r}\left(\frac{v_{\theta}}{r}\right)+\frac{1}{2r}\frac{\partial v_r}{\partial \theta}=
\frac{1}{\kappa\,M}\left[ \frac{\sin \lambda}{2}C_2- \left(\frac{\sin \lambda}{2} + \frac{1}{2\sin \lambda} \right)C_1  \right]\,\xi^{-3/2}\,\,,
\label{eq_Lrtheta}
\end{eqnarray}
where $C_1\equiv\gamma\,b(1-b)^{(\gamma-1)}$, $C_2\equiv(1-b)^{\gamma}$ and $\kappa \equiv 1/4.7$ are constants used to simplify the expressions. We use the following relationships to calculate the components of $\vec{\Lambda}$ on the surface of the disk and set $r_f=r$, as different streamlines are anchored to different radii in the disk
\begin{equation} \label{eq:vp_vr}
v_r = v_p \cos\lambda \approx v_p
\end{equation}
\begin{equation} \label{eq:vr2}
\frac{v_r}{r}=(1-b)^{\gamma}\frac{v_{\infty}}{r} = 
\frac{C_2}  {\kappa}\xi^{-1/2}r^{-1}  = \frac{C_2}{\kappa\, M}\,\xi^{-3/2} \,\,,
\end{equation}
\begin{equation}\label{eq:difvr}
\frac{\partial v_r}{\partial r}=\frac{\partial v_p}{\partial r}=
\gamma\,b(1-b)^{(\gamma-1)}\frac{v_{\infty}}{r}= 
\frac{C_1}{\kappa\,M}\,\xi^{-3/2} \,\,,
\end{equation}
\begin{equation} \label{eq:r2theta}
\frac{\partial}{\partial \theta}=
-\frac{r}{\sin\lambda}\frac{\partial}{\partial r} \,\,,
\end{equation}
where $v_\infty = \xi^{-1/2}/\kappa$, as defined above. Note that the factor of $M$ in the denominator of equations~\ref{eq_Lrr}--\ref{eq_Lrtheta} arises from the conversion of the disk radius into geometric units. It follows that $\Lambda$, and consequently $Q$, are expressed in geometric units of time, $M^{-1}$, as noted in Section~\ref{sec:dwmodel}.

%%%%%%%%%%%%%%%%%%%%%%%%%%%%%%%%%%%%%%%%%%%%%%%%%%%%%
%%% FIGURE 15
%%%%%%%%%%%%%%%%%%%%%%%%%%%%%%%%%%%%%%%%%%%%%%%%%%%%%
\begin{figure*}[t]
\centering
\includegraphics[width=1.0\textwidth, clip=true]{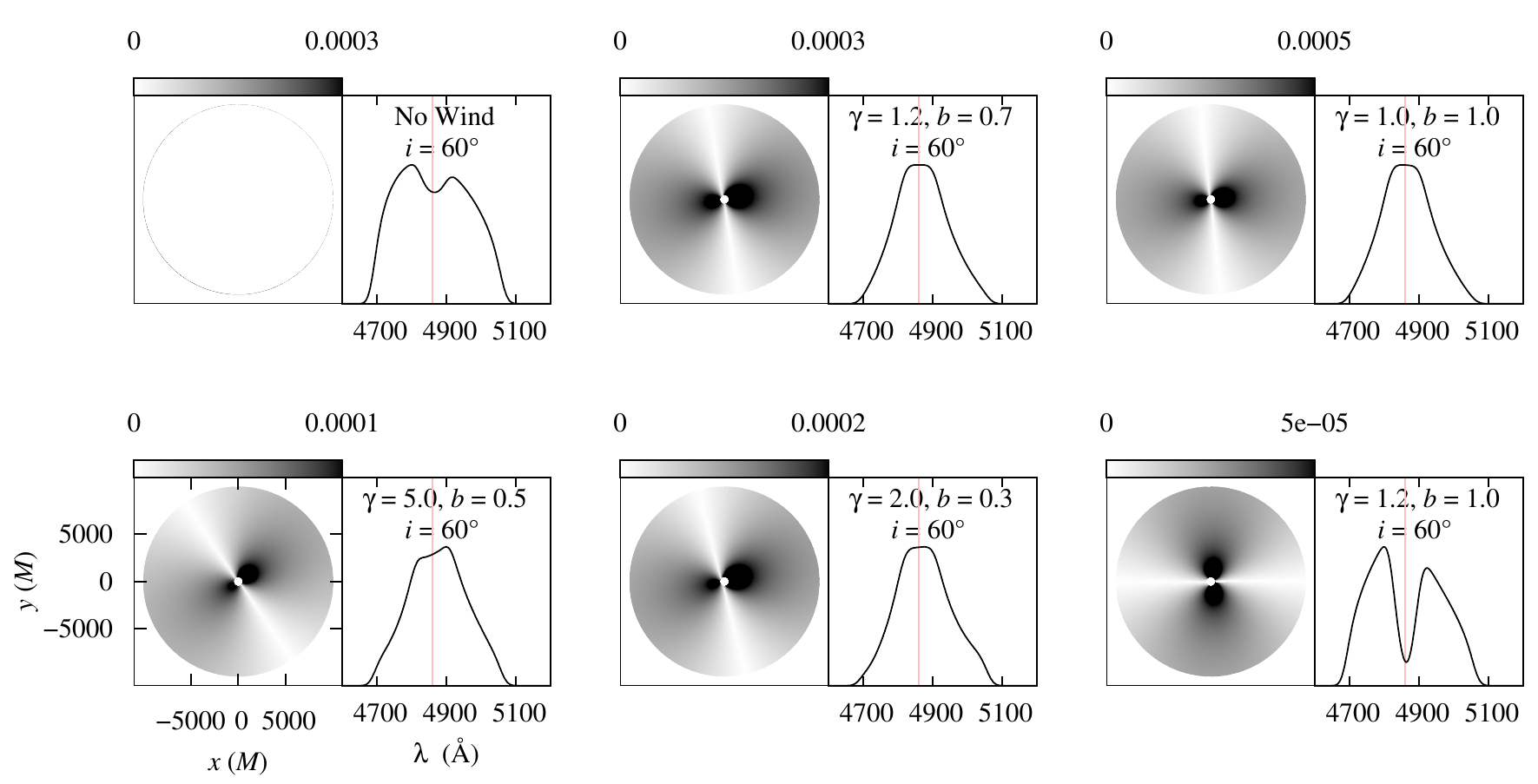}
\caption{Maps of the velocity gradient along the line of sight, $Q$, shown for different realizations of the velocity field of the wind, parametrized in terms of $b$ and $\gamma$, and for two different orientations of the observer's line of sight, $i$. The observer is located on the right and above the page, at an angle $i$ relative to the normal to the page. The rotation of the disk is counter-clockwise. Each panel also shows the resulting H$\beta$ emission-line profile associated with a given disk wind configuration. Pink vertical line at $4860.09$\AA\, marks the rest wavelength of the H$\beta$ emission line.}
\label{fig:bgamma}
\end{figure*}

Figure~\ref{fig:bgamma} shows the maps of the velocity gradient along the line of sight for different realizations of the velocity field of the wind (encoded in parameters $b$ and $\gamma$). Each panel also shows the resulting emission line (here shown as the H$\alpha$ profile) associated with a given disk wind configuration. Note that because $\tau\propto Q^{-1}$, the larger values of $Q$ correspond to the regions of lower optical depth in the disk and vice versa. The top left panel shows a double-peaked emission-line profile from a disk with no accretion disk wind, as seen by a distant observer placed on the horizontal axis extending to the right of the disk to infinity, at inclination $i=60^\circ$. The top middle panel shows the map corresponding to the choices of $b$ and $\gamma$ adopted in this work. Note that the regions of highest optical depth are expected to coincide with the regions in the disk that give rise to the largest Doppler shifts of emitted photons, as seen by a distant observer. As a consequence, these photons are missing from the profile, giving rise a narrower, single-peaked emission-line. 

The top right panel of Figure~\ref{fig:bgamma} shows the velocity gradient map for the choice of $b$ and $\gamma$ adopted by \citet{cm96} and \citet{flohic12}, which result in the emission-line profiles effectively indistinguishable from those in our model. The left and center panels in the bottom row illustrate the appearance of the velocity gradient map for the arbitrary values of $b$ and $\gamma$, considerably different from those inferred in AGNs. Finally, the bottom right panel shows the resulting map and profile calculated for a combination of $\gamma \neq 1$ and $b=1$ that violates the assumptions used in the Sobolev approximation (see the last paragraph of the previous section). 

\subsection{A.3. Density profile of the wind}\label{sec:DWdensity}

We describe the density of the wind as a decreasing function of radius from the central SBH, $\rho \propto r^{-\eta}$, where $\eta > 0$ is the density index. In order to understand the range of plausible values for $\eta$ we consider two different disk wind geometries in the context of the continuity equation
\begin{equation}
\frac{\partial \rho}{\partial t} + \nabla\cdot(\rho\vec{v})=0 \;\;.
\end{equation}
For a steady state, azimuthally symmetric wind the continuity equation must satisfy conditions $\partial/\partial t=0$ and $\partial/\partial\varphi=0$. In spherical coordinates, this gives 
\begin{equation}\label{eq:continuity}
\frac{1}{r^2}\frac{\partial \left(r^2 \rho v_r\right)}{\partial r} + \frac{1}{r\sin \theta}\frac{\partial \left(\rho v_{\theta}\sin \theta \right)}{\partial \theta} =0 \;\;.
\end{equation}
This equation can be further simplified if one considers a disk wind emerging from the emitting layer with $\theta=90^{\circ}$
\begin{equation}\label{eq_cont2}
\frac{\partial \left( \rho v_r\right)}{\partial r} + \frac{2 \rho v_r}{r} + \frac{1}{r}\frac{\partial \left( \rho v_{\theta}\right) }{\partial \theta}=0\;\;.
\end{equation}
Using the definition of $v_\theta$ from equation~\ref{eq:vtheta}, as well as \ref{eq:r2theta}, and assuming a small  opening angle of the wind streamlines, $\sin\lambda \approx \lambda \propto 1/r$, the third term in the above equation can be approximated as
\begin{equation}
\frac{1}{r}\frac{\partial \left( \rho v_{\theta}\right) }{\partial \theta} \approx \frac{\partial \left( \rho v_r\right)}{\partial r} - \frac{ \rho v_r}{r} \;\;.
\end{equation}
Using this in equation~\ref{eq_cont2} gives
\begin{equation}
2\frac{\partial \left(\rho v_r\right)}{\partial r} +\frac{\left(\rho v_r\right)}{r} =0
\end{equation}
and consequently $\left(\rho v_r\right) \propto r^{-1/2} $. Since $v_r(r=r_f) \propto v_{\infty} \propto r^{-1/2}$ it follows that the wind mass density must be constant as a function of radius, $\rho\propto r^0$. Therefore, a steady state, azimuthally symmetric wind, characterized by a small opening angle, is well described by the density index $\eta = 0$.

Repeating the same exercise for a steady state, spherically symmetric wind, one obtains $\eta=1.5$. In this case the continuity  equation~\ref{eq:continuity} reduces to:
\begin{equation}
\frac{\partial \left(r^2 \rho v_r\right)}{\partial r}=0
\end{equation}
which yields $\left(\rho v_r\right) \propto r^{-2}$ and hence $\rho\propto r^{-1.5}$. In this work we choose $\eta=1$ as a representative intermediate value.

%%%%%%%%%%%%%%%%%%%%%%%%%%%%%%%%%%%%%%%%%%%%%%%%%%%%%
%%% FIGURE 16
%%%%%%%%%%%%%%%%%%%%%%%%%%%%%%%%%%%%%%%%%%%%%%%%%%%%%
\begin{figure}[t]
\centering
\includegraphics[width=0.39\textwidth, clip=true]{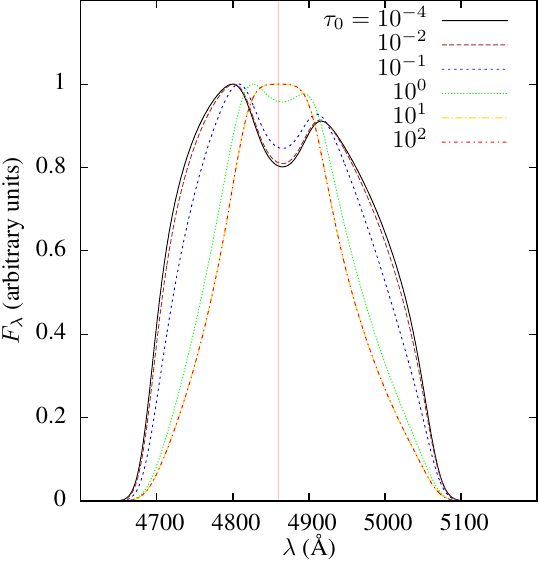}
\caption{H$\beta$ broad emission-line profiles from a circular disk calculated for different values of the disk wind optical depth, $\tau_0$: $10^{-4}$ (black, solid), $10^{-2}$ (maroon, long-dash), $10^{-1}$ (blue, dashed),  $1$ (green, dotted), $10$ (yellow, long-dash-dot) and $10^2$ (red, dash-dot). Pink vertical line at $4860.09$\AA\, marks the rest wavelength of the H$\beta$ emission line. The parameters used in calculation of the profiles are: $\xi_{\rm in}=500 \, M$, $\xi_{\rm out}=10,000 \, M$, $i=60^{\circ}$, $\sigma=850 \,{\rm km\,s^{-1}}$, $\epsilon \propto r^{-3} $, $\gamma=1.2$, $b=0.7$, $\eta=1$, and $\lambda(R_{\rm in}) = 10^\circ$.}
\label{fig:tau}
\end{figure}

\subsection{A.4. Optical depth of the wind}\label{sec:DWtau}

The final ingredient in the calculation of the optical depth of the disk wind is the normalization factor, $\tau_0 = \varkappa \,\rho_0 \,\sigma$, used in equation~\ref{eq:depth1}. In this form $\tau_0$ is used to define the optical depth of the surface layer at the inner edge of the BLR, in the direction perpendicular to the disk plane, so that $\tau \approx 5\tau_0$ ($7\tau_0$) at $R_{{\rm in}, i} = 500\,M_i$ ($1000\,M_i$). It is worth emphasizing that $\tau_0$ dimensionally represents optical depth scaled by the mass of the disk's central SBH, $\propto M_i^{-1}$. Combined with the velocity gradient of the wind along the line of sight, Q, which exhibits the same dependence, it results in the dimensionless parameter $\tau$. 

As noted in \S\,\ref{sec:dwmodel}, the emission line profiles in this work are calculated for a range of optical depths, given by $\tau_0 = 10^{-4}, 0.1, 1, 100$. In addition to these Paper~I presents emission-line profiles for $\tau_0 = 0$, for a model without the disk wind. Because the profiles calculated for $\tau_0 = 0$ and $10^{-4}$ are very similar, we use them interchangeably. Moreover, we verify that profile shapes do not change in shape once $\tau_0 > 100$ and do not explore the values of optical depth beyond this threshold. Such high line optical depths effectively imply a weak profile that would be difficult to discern in real spectra, due to the low contrast with respect to the continuum and the presence of noise.

Figure~\ref{fig:tau} illustrates the transition from the broader, double-peaked to the narrower, single-peaked emission-line profiles with increasing optical depth. They are included for comparison with the previously published works, as well as an intermediate verification step for those wishing to reproduce this calculation.  The profiles are from a circular disk, illuminated by a single, central AGN and calculated for values of $\tau_0$ and other parameters used in this work. As discussed in \S\,\ref{sec:NOP} and \S\,\ref{sec:char}, a similar trend is also reflected in the composite profiles calculated from triple disk configurations of BLRs in circumbinary regions.

\subsection{A.5. Comparison of modeled emission-line profiles to the observed profiles}\label{sec:compare}

This test is not straightforward to carry out due to several differences in the way the synthetic and observed samples were constructed. Firstly, in the synthetic sample we choose uniform distributions of the SBHB parameters (shown in Table~\ref{table:parameters}) in order to obtain a uniform but not necessarily dense coverage of the SBHB parameter space. We prefer this agnostic approach to modeling because the distribution functions for various SBHB parameters are still not well constrained. Something that can be expected with a reasonable level of confidence however is that the mass ratios and orbital separations of the observed set of SBHBs are not uniformly distributed. Moreover, the fact that our database has finite size inevitably means that we are not capturing all profile shapes that an SBHBs in circumbinary disks might have. Consequently, the statistical distribution functions for simulated profiles in Figures~\ref{fig:catAIPPS} -- \ref{fig:catAIPCS} and those for the observed sample shown in Figures~\ref{fig:obsAIPPS} -- \ref{fig:obsAIPCS}, may share some subset of profiles but they are not drawn from the same parent distribution, and in the case of the synthetic profiles, not in a random way.

Secondly, the observed SBHB candidates have been selected based on a few more criteria, which we do not apply to the modeled profiles. Specifically, the SBHB candidates selected for further monitoring by E12 have the broad component of the emission-line profile offset by $\gtrsim 1000\,{\rm km\,s^{-1}}$, a practical requirement that makes it easier to separate the broad and narrow components of the line during the analysis. Our database however includes profiles with all values of velocity offsets, including those with $< 1000\,{\rm km\,s^{-1}}$. Similarly, E12 only choose for followup candidates with the H$\beta$ emission-line profiles which in the first epoch of observation appear single-peaked {\it before} subtraction of the narrow H$\beta$ and [O\,\textsc{iii}] lines. Since we only model the emission from the BLR, we cannot predict the appearance of the profiles with superimposed narrow-line emission. We instead compare our modeled profiles directly to the broad-line component of the observed H$\beta$ profiles \citep[see][for discussion of a method for subtraction of the narrow lines.]{runnoe15}. These differences preclude us from carrying out an apples-to-apples comparison of the modeled and observed line profiles at this point.
 
 We can nevertheless perform a relative comparison by asking: are the modeled profiles more similar to the profiles of SBHB candidates or the control sample of regular AGN? For this test we design a following simple statistic

\begin{equation}\label{eq:SD}
\mathcal{R}_{x-y}=\frac{\sum_{i} \sum_{j} \, \left(P_{\rm SBHB}(x_i, y_j)-P_{\rm mod}(x_i, y_j)\right)^2 }{ \sum_{i} \sum_{j} \, \left(P_{\rm AGN}(x_i, y_j)-P_{\rm mod}(x_i, y_j)\right)^2}
\end{equation}
where $P_{\rm mod}$, $P_{\rm SBHB}$, and $P_{\rm AGN}$ represent the two-dimensional probability distributions of profiles drawn from the modeled, the SBHB candidate, and the regular AGN sample, respectively. Each distribution represents the probability of finding a profile in a given ``pixel" of a discretized two-dimensional map, for which  coordinates ($x_i$, $y_j$) mark the center of the pixel with the width $\Delta x$ and $\Delta y$. For the purpose of this comparison we choose the size of each pixel to be a hundredth of the full numerical range of a distribution, i.e. $\Delta x = \left( x_{\rm max} - x_{\rm min} \right)/100$.  We have confirmed that this choice does not affect the value of $\mathcal{R}_{x-y}$ by comparing the smoothed distributions of the observed and modeled profiles for a variety of grid sizes, ranging from  $25\times25$ to $1000\times1000$. Defined in this way $\mathcal{R}_{x-y}$ becomes a simple test of similarity in the overall shape of the two-dimensional distributions, without any a priory assumptions about their origin.

%%%%%%%%%%%%%%%%%%%%%%%%%%%%%%%%%%%%%%%%%
%%%  TABLE 2
%%%%%%%%%%%%%%%%%%%%%%%%%%%%%%%%%%%%%%%%%
\begin{deluxetable}{clcl}[t]  
\tabletypesize{\normalsize}
\tablecolumns{4}
\tablewidth{0pt} 
\tablecaption{Comparison of observed and modeled datasets\label{table:ratio}}
\tablehead{$x-y$ & $\mathcal{R}_{x-y}$ & $x-y$ & $\mathcal{R}_{x-y}$}
\startdata
AIP-PS*   & 0.20  & KI-CS   & 0.55 \\
FWHM-C   & 0.32 & AIP-CS*    & 0.63\\
FWQM-CS* & 0.34 & KI-$\sigma_2$ & 0.79 \\
FWHM-AI  & 0.40 & KI-AI    & 0.90\\
FWHM-AIP  & 0.45 & AIP-$\sigma_2$ & 0.94\
\enddata
\tablecomments{AIP - Pearson skewness coefficient. PS - Peak shift. FWHM, FWQM - Full width at half and quarter maximum, respectively. C - Location of the centroid. CS - Centroid shift. AI - Asymmetry index. KI - Kurtosis index. $\sigma_2$ - Second moment. * Marks distributions presented in this work.}
\label{table:ratio}
\end{deluxetable}

Table~\ref{table:ratio} shows $\mathcal{R}_{x-y}$ values calculated for different profile shape parameters commonly used to analyze the emission-line profiles (see Paper~I for their definitions). The synthetic profiles used in calculation of $\mathcal{R}_{x-y}$ are drawn from the model C-3DW-100.  In order to mimic the selection process of the E12 search as closely as possible, before calculating $P_{\rm mod}$ we remove profiles with peak shifts $<1000\,{\rm km\,s^{-1}}$ and those that have more than one peak from the modeled sample. In this statistic, $\mathcal{R}_{x-y} = 0$ when the probability distribution of the SBHB candidates is precisely matching that of the modeled profiles. The values of $\mathcal{R}_{x-y} <1$ correspond to a higher degree of similarity between the modeled and SBHB candidate profiles than the modeled and regular AGN profiles. Conversely, $\mathcal{R}_{x-y} >1$ indicates a higher degree of similarity between the modeled and regular AGN profiles than the modeled and candidate SBHB profiles.

As it can be seen from Table~\ref{table:ratio}, $\mathcal{R}_{x-y} <1$ for all statistical distribution pairs, where we mark with an asterisk the pairs presented in Figures throughout this work. It is worth noting that the distributions that we have a priori identified as more robust, because they measure the dominant features in the bulk of the profile (AIP, PS, CS similar to C, FWHM similar to FWQM), favor similarity between the observed SBHB candidate and modeled profiles. On the other hand, higher moments of the profile flux distribution, which tend to be more easily affected by the noise ($\sigma_2$, AI, KI), on average result in higher values of $\mathcal{R}_{x-y}$. This simple comparison therefore seems to support the hypothesis that the shapes of the emission-line profiles from a sample of observed SBHB candidates are more consistent with the binary model than are regular AGNs. 

A related pertinent question is: {\it how large an observed SBHB sample should be for a meaningful statistical comparison between the observed and synthetic profiles?} Since 2D profile distributions, like AIP-PS shown in the top left panel of Figure~\ref{fig:catAIPPS}, are described by their shape and dynamic range (illustrated by color), we want to find the minimum number of profiles needed to represent the shape and colors of this distribution. The AIP-PS distribution shown in the top left panel of Figure~\ref{fig:catAIPPS} has $N=12$ million profiles and spans a dynamic range of 6 orders of magnitude. We use the Cochran's formula \citep{cochran77} to estimate a sample size large enough so that there is a finite number of profiles within the yellow (or red, or green) color region of the figure with 95\% confidence. 
\begin{eqnarray}
n=\frac{N n_0}{N+n_0-1} \;\;\;\; {\rm and}\;\;\;\;n_0=\frac{Z^2 P (1-P)} {e^2}\,\,.
\end{eqnarray}
Here $n$ and $N$ are the sizes of the sample and the entire population, respectively. The parameter $n_0$ asymptotes to $n$ in the limit when $N$ approaches infinity.  $P$ is a proportion of the profiles with the relevant attribute (color) in the population. Since we want to draw profiles that represent the yellow color region of the distribution, we estimate $P_{\rm yellow}\approx 10^{2.5}/10^6=10^{-3.5}$, where the exponents $2.5$ and $6$ correspond to the values on the color bar. $Z$ represents the number of standard deviations away from the expected proportion $P_{\rm yellow}$ in a normal distribution (also often referred to as ``sigma"). For example, $95\%$ confidence level  corresponds to $Z=1.96 \approx 2$. The parameter $e$ determines the margin of error, so that the sample will have some fraction of profiles in the yellow region, within the range $\left[P_{\rm yellow} \pm e \right]$. Since the value of $P_{\rm{yellow}}$ is small, we can choose $e=P_{\rm yellow}$, so that the fraction of profiles in the sample drawn from the yellow region is in the range from 0 to $2P_{\rm{yellow}}$.

With these values, we can estimate that the sample size of profiles, required to meaningfully reproduce the distribution of profiles in Figure~\ref{fig:catAIPPS}, down to at least the green, yellow, red contour region, is approximately $n_{\rm green}\sim 10^3$, $n_{\rm yellow}\sim 10^4$ and $n_{\rm red}\sim 10^5$, respectively. In the context of comparisons carried out in this work, these numbers can be interpreted as the minimum number of observed spectra necessary to compare the distributions of the observed and synthetic profiles. As described in paragraph~2 of Section~\ref{sec:implications}, at the time of this analysis we used 330 spectra of SBHB candidates, and 527 spectra of control sample AGNs. These numbers indicate that at present time a comparison can be made at the level of the blue and green contour regions in Figure~\ref{fig:catAIPPS}, with a confidence level of only $Z\approx 1$ (or about $60-70$\%). Future surveys are however expected to increase the number of AGN spectra (and spectra time series) by $1-2$ orders of magnitude, so we anticipate that our models will be useful beyond the comparison with the current data.
Similar analysis applies to the remaining profile distributions. We defer a more detailed comparison of individual profiles of SBHB candidates with the synthetic database to Paper~III of the series.

%%%%%%%%%%%%%%%%%%%%%%%%%%%%%%%%%%%%%%%%%%%%%%%%%%%%%
%%% REFERENCES
%%%%%%%%%%%%%%%%%%%%%%%%%%%%%%%%%%%%%%%%%%%%%%%%%%%%%
\bibliographystyle{apj}
\bibliography{apj-jour,smbh}

\end{document}